 \documentclass[twocolumn,trackchanges]{aastex61}

\hypersetup{linkcolor=blue,citecolor=blue,filecolor=blue,urlcolor=blue}
\shorttitle{UDGs in the HCG\,95 Field}
\shortauthors{Shi et al.}

\begin{document}

\title{Deep Imaging of the HCG 95 Field.\,I.\,Ultra-diffuse Galaxies}

\correspondingauthor{Xian~Zhong~Zheng}
\email{ddshi@pmo.ac.cn, xzzheng@pmo.ac.cn}

\author[0000-0002-0786-7307]{Dong~Dong~Shi}
\affil{Purple Mountain Observatory, Chinese Academy of Sciences, 2 West Beijing Road, Nanjing 210008, China}
\affil{School of Astronomy and Space Sciences, University of Science and Technology of China, Hefei 230026,China}

\author{Xian~Zhong~Zheng}
\affiliation{Purple Mountain Observatory, Chinese Academy of Sciences, 2 West Beijing Road, Nanjing 210008, China}
\affiliation{School of Astronomy and Space Sciences, University of Science and Technology of China, Hefei 230026,China}
\affiliation{Chinese Academy of Science South America Center for Astronomy, China-Chile Joint Center for Astronomy, Camino EI Observatorio $\#1515$, Las Condes, Santiago, Chile}

\author{Hai~Bin~Zhao}
\affiliation{Purple Mountain Observatory, Chinese Academy of Sciences, 2 West Beijing Road, Nanjing 210008, China}
\affiliation{Key Laboratory of Planetary Sciences, Chinese Academy of Sciences, 2 West Beijing Road, Nanjing 210008, China}


\author{Zhi~Zheng~Pan}
\affiliation{Purple Mountain Observatory, Chinese Academy of Sciences, 2 West Beijing Road, Nanjing 210008, China}

\author{Bin~Li}
\affiliation{Purple Mountain Observatory, Chinese Academy of Sciences, 2 West Beijing Road, Nanjing 210008, China}
\affiliation{Key Laboratory of Planetary Sciences, Chinese Academy of Sciences, 2 West Beijing Road, Nanjing 210008, China}

\author{Hu~Zou}
\affiliation{Key Laboratory of Optical Astronomy, National Astronomical Observatories, Chinese Academy of Sciences, Beijing 100012, China}

\author{Xu~Zhou}
\affiliation{Key Laboratory of Optical Astronomy, National Astronomical Observatories, Chinese Academy of Sciences, Beijing 100012, China}

\author{KeXin~Guo}
\affiliation{Kavli Institute for Astronomy and Astrophysics, Peking University, Beijing 100871, China}

\author{Fang~Xia~An}
\affiliation{Purple Mountain Observatory, Chinese Academy of Sciences, 2 West Beijing Road, Nanjing 210008, China}

\author{Yu~Bin~Li}
\affiliation{Yunnan Observatories, Chinese Academy of Sciences, 396 Yangfangwang, Guandu District, Kunming 650216, China}
\affiliation{Center for Astronomical Mega-Science, Chinese Academy of Sciences, 20A Datun Road, Chaoyang District, Beijing 100012, China}
\affiliation{Key Laboratory for the Structure and Evolution of Celestial Objects, Chinese Academy of Sciences, 396 Yangfangwang, Guandu District, Kunming 650216, China}





\begin{abstract}
We present a detection of 89 candidates of ultra-diffuse galaxies (UDGs) in a 4.9\,degree$^2$ field centered on the Hickson Compact Group 95 (HCG\,95) using deep $g$- and $r$-band images taken with the Chinese Near Object Survey Telescope. 
This field contains one rich galaxy cluster (Abell\,2588 at $z$=0.199) and two poor clusters (Pegasus~I at $z$=0.013 and Pegasus~II at $z$=0.040).  The 89 candidates are likely associated with the two poor clusters,  giving about 50 $-$ 60 true UDGs with a half-light radius $ r_{\rm e} > 1.5$\,kpc and a central surface brightness $\mu(g,0) > 24.0$\,mag\,arcsec$^{-2}$.  Deep $z\arcmin$-band images are available for 84 of the 89 galaxies from the Dark Energy Camera Legacy Survey (DECaLS), confirming that these galaxies have an extremely low central surface brightness. Moreover, our UDG candidates are spread over a wide range in $g-r$ color, and $\sim$26\% are as blue as normal star-forming galaxies, which is suggestive of young UDGs that are still in formation. 
Interestingly, we find that one UDG linked with HCG\,95 is a  gas-rich galaxy with H\,{\small I} mass 1.1$\times 10^{9}$\,$M_{\odot}$ detected by the Very Large Array, and has a stellar mass of $M_\star \sim 1.8 \times 10^{8}$\,$M_{\odot}$.  
This indicates that UDGs at least partially overlap with the population of nearly dark galaxies found in deep H\,{\small I} surveys.  Our results show that the high abundance of blue UDGs in the HCG\,95 field is favored by the environment of poor galaxy clusters residing in H\,{\small I}-rich large-scale structures. 
\end{abstract}

\keywords{galaxies: group: individual: (HCG\,95) --- 
galaxies: evolution --- galaxies: structure}



\section{Introduction} \label{sec:intro}

Extremely low surface brightness (LSB) galaxies with unexpectedly large sizes, namely ultra-diffuse galaxies (UDGs), have recently drawn much attention to the understanding of their formation and evolution. Findings of dwarf galaxies with extreme LSB have been reported two decades ago \citep[e.g.,][]{Impey1988,Bothun1991}. UDGs were detected in the Coma cluster using deep low-resolution imaging data obtained with small telescopes \citep{van Dokkum2015a}, and more UDGs were found in nearby galaxy clusters since then \citep{Koda2015,Mihos2015,Munoz2015,Beasley2016,vanderBurg2016,Koch2016,Roman2017}. Differing from the classical LSB galaxies that usually have a central surface brightness down to $\mu(B,0) \sim 22-23$\,mag\,arcsec$^{-2}$ \citep{Impey1997, Impey2001, Ceccarelli2012,Geller2012}, UDGs have a much lower central surface brightness ($\mu(g,0)=24-26$\,mag\,arcsec$^{-2}$), but their half-light radius $r_{\rm e} >1.5$\,kpc is comparable to that of typical $L^\ast$ galaxies, and their stellar mass ($<\sim 10^{8}$\,$M_{\odot}$) is two orders of magnitude lower \citep{van Dokkum2015a}. They usually appear to be red, relatively round, and morphologically featureless in galaxy clusters, but are blue and irregular in the field \citep{Merritt2016,Leisman2017,Roman2017,RomanTrujillo2016,Trujillo2017}. The ultra-diffuse nature of UDGs remains little explored. 

Increasing effort has been made to understand the formation of UDGs and their connections with other galaxy populations. However, the observations have so far led to diverse conclusions. Measurements of velocity dispersion of some UDGs (e.g., Dragonfly~44 in Coma) suggest that they are likely to be failed $L^\ast$ galaxies that are overwhelmingly dominated by dark matter, although the reasons for their extremely low star formation efficiency in the past are still unknown \citep{van Dokkum2016}. Recently, \cite{van Dokkum2017} reported that the halo masses of largest Coma UDGs (Dragonfly~44 and Dragonfly~X1) are at the level of $\sim 5 \times 10^{11}\,M_{\odot}$, lower than the previous estimate. The evidence of a high abundance of globular clusters in UDG Dragonfly~17 supports that it could be a failed galaxy with a halo mass similar to the halo masses of the Large Magellanic Cloud (LMC) or M33 \citep{BeasleTrujillo2016,Peng2016}. Additionally, VCC~1287, a UDG that is found in the Virgo cluster and with a halo mass estimated from the kinematics of globular clusters, is thought to belong to the dwarf galaxy population, although it is rather extended ($r_{\rm e}=2.4$\,kpc) \citep{Beasley2016}. More observations for a large sample of UDGs are thus needed to study the properties of UDGs in detail.

From the theoretical perspective, \cite{Amorisco2016} suggested that UDGs are part of the dwarf galaxy population, but with extreme spin properties. Strong outflows are proposed to be responsible for the termination of star formation and gas cooling at early cosmic time and for the formation of UDGs seen today \citep{Janowiecki2015,Di Cintio2016}. Since UDGs have first been discovered in the Coma cluster, the processes confined to dense environment such as ram pressure and tidal stripping are likely important mechanisms for generating such galaxies. Recent progresses have reported the discovery of UDGs in various environments, not only in clusters. Of these works, \citet{Smith2016} reported two UDGs in a compact group (HCG~44), which is  less dense than massive galaxy clusters. In addition, UDGs are found in even sparser environments, such as the vicinity of M101 \citep{Merritt2016} and poor galaxy groups \citep{RomanTrujillo2016}. These works together demonstrate that UDGs do not exist exclusively in the cluster environments. 

Hickson compact groups (HCGs) have a high density that is comparable to the center of clusters, and they have a low velocity dispersion similar to those of loose groups ($<\sigma>\,\sim\,200\,\rm km\,s^{-1}$), providing a unique environment to sustain galaxy evolution \citep{Hickson1992}. Members of HCGs may almost undergo continuous gravitational perturbations and show signs of enhanced activities such as like bluer optical colors or higher radio continuum power in the violently interacting galaxies and close pairs \citep{Verdes-Montenegro1997}. In addition, many HCGs contain a diffuse background light envelope in observations, and these diffuse envelopes may be stripped material from cluster member galaxies \citep{DaRocha2005}.  HCGs are also suggested to be favorable places for searching LSB galaxies and UDGs \citep{Ordenes-Briceno2016}.

We carry out a search for extreme LSB galaxies (including UDGs) in a field centered on HCG\,95. Our observations cover a field of 3$^{\circ}$\,$\times$\,3$^\circ$. In this paper, we present 89 candidates of extreme LSB galaxies (with 57 plausible UDGs) found in deep $g$- and $r$-band imaging and their spatial distribution in the HCG\,95 field. 
In Section~\ref{sec:observe} we describe the observation and data processing. UDG candidates are selected and their properties are given in Section~\ref{sec:candidates}, and in Section~\ref{sec:discussion} we discuss our findings. Finally, we summarize our results in Section~\ref{sec:summary}. We adopt an HCG 95 distance modulus of 36.05\,mag (162.2\,Mpc) with the following cosmology: $\Omega_{\rm M} = 0.3$, $\Omega_{\rm \Lambda} =0.7$ and $ H\rm_0 = 70$\,km\,s$^{-1}$\,Mpc$^{-1}$.  We use the AB magnitude system throughout this work.

\section{Observation and Data Reduction} \label{sec:observe}

Our imaging observations of a 3$\degr \times 3\degr$ field centered on HCG\,95 were taken with the Chinese Near Object Survey Telescope (CNEOST) of diameter 1.04/1.20 m in Xuyi Station \citep{Zhang2013,Zhang2014}. The telescope is equipped with a 10k$\times$10k STA\,1600 CCD with 16 readout channels, providing a field of view (FOV) of 3$\degr \times  3\degr$ and a pixel scale of 1$\farcs$029. At the distance of HCG\,95 (162.2\,Mpc), this corresponds to a resolution of 0.784\,kpc per pixel. The Sloan Digital Sky Survey (SDSS) $g$ and $r$ filters were chosen for the imaging observations over five nights from 2015 October 10 to 16. Each exposure takes 90\,s. The integration time is 16.15 hr and 10.65 hr for $g$ and $r$, respectively.  Before the starting and after the end of every night, we obtained bias and dark images as well as sky flats throughout each filter.

\begin{figure*}
 \begin{center}
  \includegraphics[height=0.7\textwidth]{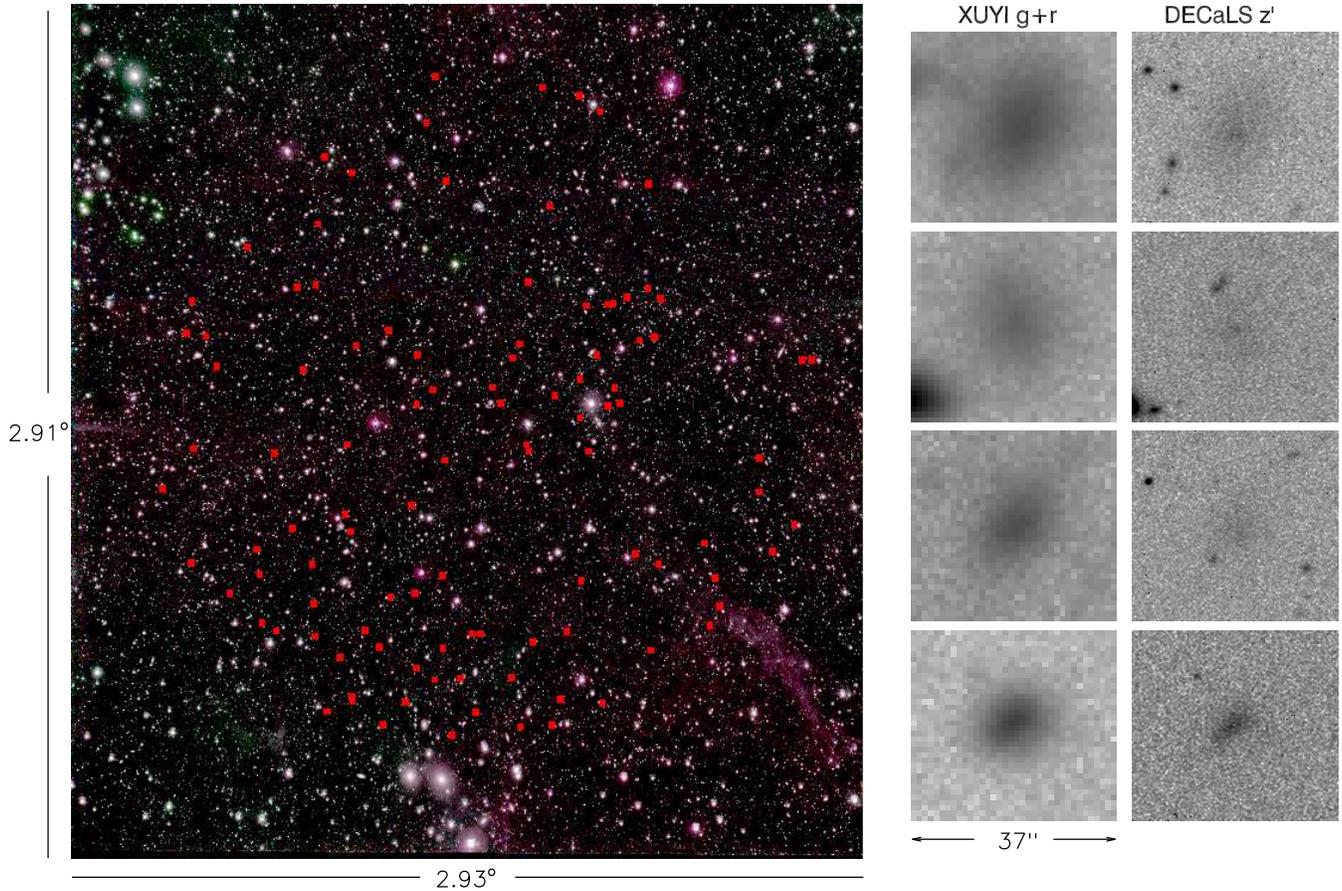}
  \caption{Left: the deep $g+r$ color image of the HCG\,95 field covering a sky area of $2\fdg93 \times 2\fdg91$. Red squares mark the 105 UDG candidates. Right: typical UDGs viewed in the images from Xuyi and the Dark Energy Camera Legacy Survey (DECaLS). The size of the postage stamp images is $37\arcsec \times 37\arcsec$.}
  \label{fig:fig1}
 \end{center}
\end{figure*}

\begin{figure*}
 \setlength{\abovecaptionskip}{5pt}
 \begin{center}
  \includegraphics[trim=6mm 5mm 0mm 25mm,height=0.65\textwidth]{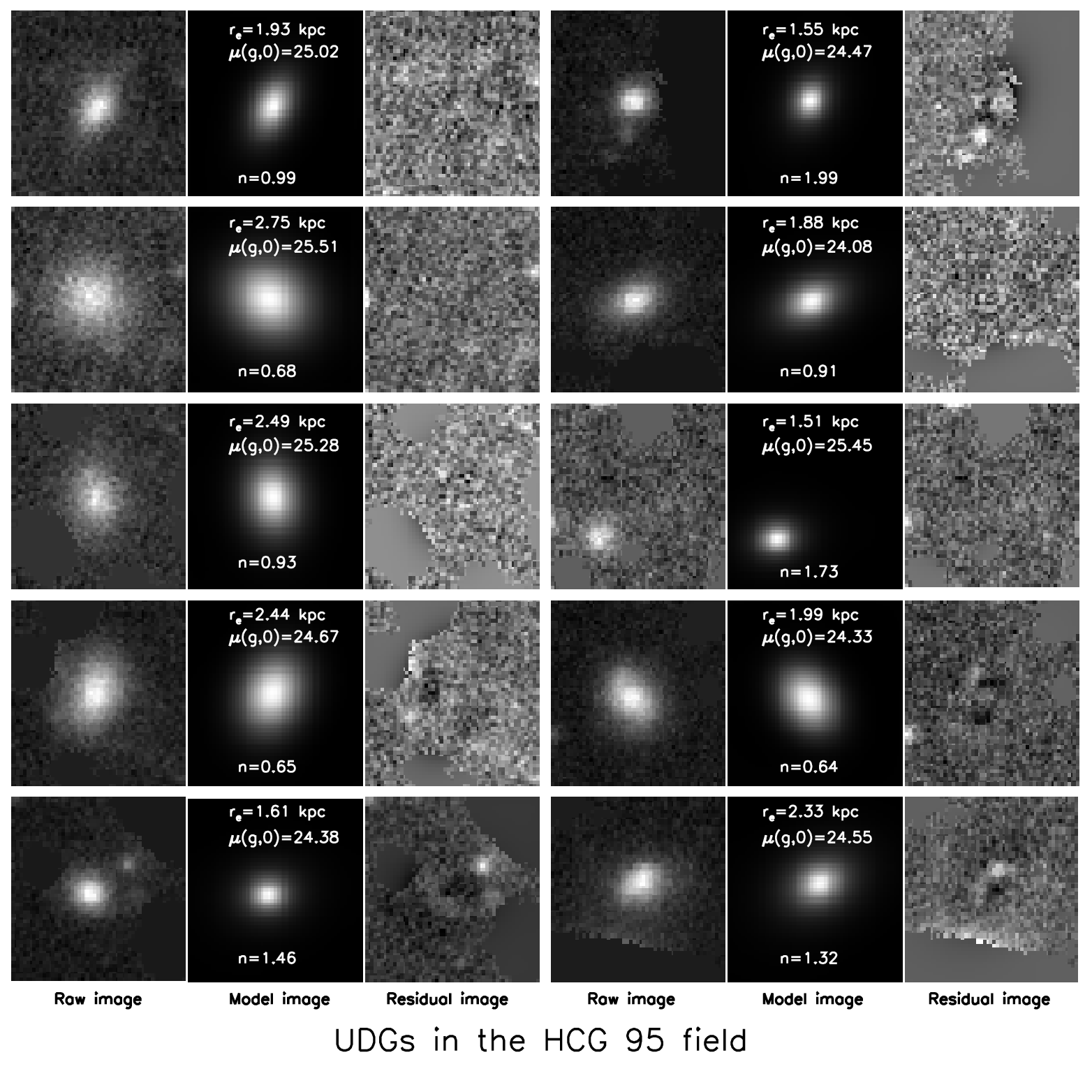}
  \caption{Ten examples of GALFIT models to UDG images. The three columns from left to right show the $g$-band image, the GALFIT model, and the residual image. The best-fit S{\'e}rsic profiles are described by the half-light radius $r_{\rm e}$ (kpc), the central surface brightness $\mu(g,0)$ (in units of mag\,arcsec$^{-2}$), and the S{\'e}rsic index $n$. The size of every postage stamp image is $56\arcsec \times 56\arcsec$.}
  \label{fig:fig2}
 \end{center}
\end{figure*}

The raw image data were reduced following standard procedures, including bias and dark subtraction, flat-fielding, and removal of bad/hot pixels. Each exposure image contains 10k$\times$10k pixels divided into 16 regions to be read out. The effective gain of the readout channels differs slightly from each other. Sky flat frames are used to determine the effective gains of the readout channels. The raw images were corrected to have the same effective gain before flat-fielding. For the sky flat frames, we normalized each to the median of the central 6k$\times$6k region and then divided by the best-fit slant plane to the same region in order to correct for the inhomogeneity within a single frame. Then these flat frames were combined together to generate the final flat. Again, the final flat was divided by the best-fit slant plane determined from the central  6k$\times$6k region in order to remove systematic structures.  The final flat frame is used for flat-fielding that removes pixel-to-pixel variations. Bad or hot pixels were identified with values below 0.9 or above 1.1 in the final flat frame. Perfect flat-fielding could improve the detection depth for faint structure because the detection limit is constrained by systematic errors, which are dominated by flat-fielding \citep{Abraham2014}.         
The vignetting effect becomes significant at the edges of a wide-field image. This effect can be mostly corrected for through flat-fielding.

A catalog of 12,922 stars in the HCG\,95 field from SDSS is used as the reference to determine astrometric distortions and calibration.   Source detection and photometry was done using the software SExtractor \citep{BertinArnouts1996}, and SCAMP \citep{Bertin2006} was used to compute an astrometric solution for individual science images, giving a typical accuracy of RMS$=0\farcs 1$ in astrometry. Only bright stars of $14 < g < 18$\,mag with photometry from SDSS are adopted as the reference sources for photometric calibration.  All science images were calibrated  to have a constant photometric zero point $zp$ = 25.0\,mag.  We fit a slant plane to the background and subtracted it from each image. This helps to reduce systematic errors in image combination.  Weight images were generated to mask bad/hot pixels and saturated pixels.  Cross-talks caused by saturated stars were removed following \cite{Freyhammer2001}.

Finally, we selected the good-quality science images and stacked those of the same filter together using the software tool SWARP \citep{Bertin2002}, yielding the final $g$- and $r$-band science images with effective integration times of 13.15 hr and 9.525 hr, respectively. The $3\,\sigma$ depth of the two images (in a $ 10$\arcsec$ \times 10$\arcsec$ $ box) reaches $\mu(g) \sim 29.16$\,mag\,arcsec$^{-2}$ and $\mu(r) \sim 28.38$\,mag\,arcsec$^{-2}$, respectively. The depths are comparable to those given in \cite{FliriTrujillo2016} and \cite{Koda2015}. The $g + r$ combined image is shown in the left panel of Figure~\ref{fig:fig1}.

The FWHM of the point spread function (PSF) is typically 4{$\arcsec$} and 3{$\arcsec$} in the central region of the $g$- and $r$-band science image, respectively. We note that the PSF of the images obtained with CNEOST slightly increases from the center to the edges of the FOV. We use stars identified from SDSS to trace the PSF variation and build the 2-dimensional map of PSF FWHM through best fitting a third-order polynomial function to the FWHM map of the selected stars. The best-fit PSF FWHM maps in $g$ and $r$ are shown in the Appendix in Figure~\ref{fig:figA1}.

\section{UDG Candidates} \label{sec:candidates}

Our goal is to search for UDGs that have large sizes and very LSB.  We run SExtractor \citep{BertinArnouts1996} in dual mode with the configuration optimized for the detection of extended LSB objects and extract source catalogs from the stacked science images. The $g$-band image is used for source detection because it is somewhat deeper and has a slightly better resolution (see Figure~\ref{fig:figA1}). We limit the source detection to the central area of diameter $<2\fdg5$ to minimize the edge effects (such as vignetting or comatic aberration).   We obtained the $g$- and $r$-band photometric catalogs consisting of 31,230 objects. 

The widely used criteria for UDGs selection are $r_{\rm e} > 1.5$\,kpc and $\mu(g,0) > 24.0$\,mag\,arcsec$^{-2}$ \citep{van Dokkum2015a}. Given that UDGs generally obey a rather flat profile, we average the total magnitude over the area within the PSF-corrected half-light radius as the central surface brightness, following \cite{van Dokkum2015a}. We cross correlate our catalog with the photometric catalog of the same field from the SDSS with a tolerance radius of $r=1\farcs5$, and exclude stars, bright sources with $\mu(g,0) < 24.0$\,mag\,arcsec$^{-2}$, and compact sources with $r_{\rm e} < 1\farcs92$.  Here the cut $1\farcs92$ corresponds to a size of 1.5\,kpc at the distance of HCG\,95.   The objects in our catalog with stellarity parameter CLASS\_STAR$>$0.15 are classified as compact/point sources and those with a color out of $g-r<0$ or $g-r>1.2$ are treated as false sources, including noise contaminators and sparks from bright sources.  We obtained 464 SDSS-detected galaxies of  $\mu(g,0) > 24.0$\,mag\,arcsec$^{-2}$ and $r_{\rm e} > 1\farcs92$, and 1771 sources from our deep $g$ and $r$ observations without detection in the SDSS.  
For the preselected 2235 sources, we estimated the intrinsic half-light radii and central surface brightness from our deep $g$-band image to further identify UDGs. 
The size measurement for galaxies needs to correct the broadening effect by PSF. We extracted the PSF size from the 2D PSF FWHM maps shown in Figure~\ref{fig:figA1} for the objects of given positions.  We roughly estimated the intrinsic effective radius with $ r_{\rm e} = \sqrt{ r_{\rm obs}^{2}- R_{\rm psf}^{2}} > 1\farcs92$ following \cite{Bouche2015}, where $ r_{\rm obs}$ is the directly observed half-light radius, and $R_{\rm psf}$ is the half-light radius of the PSF (see Figure~\ref{fig:figA3} for more details).  
We estimated that 644 of the 2235 objects satisfy $r_{\rm e} > 1\farcs92$ and $\mu(g,0) > 24.0$\,mag\,arcsec$^{-2}$.

We realize that the preselected 644 objects that satisfy the selection criteria for UDGs may still be contaminated by blended sources. In particular, the low resolution of our deep $g$- and $r$-band images together with the source detection configured for extended sources would often mistake blended faint sources as UDG candidates. On the other hand, the degradation of the image quality toward the edges of the images increases the uncertainties in identification of UDGs.   
We visually inspect the preselected targets with the aid of SDSS DR12 Image List Tool\footnote{\url{http://skyserver.sdss.org/dr12/en/tools/chart/listinfo.aspx}} to examine their morphologies. 
Removing binary sources and other artifacts of misclassification, we obtained a sample of 105 UDG candidates over an area of 4.9\,degree$^2$ in the HCG\,95 field.

\begin{figure*}
 \begin{center}
 \setlength{\abovecaptionskip}{-40pt}
  \includegraphics[trim=15mm 0mm 0mm 0mm,clip,height=0.72\textwidth]{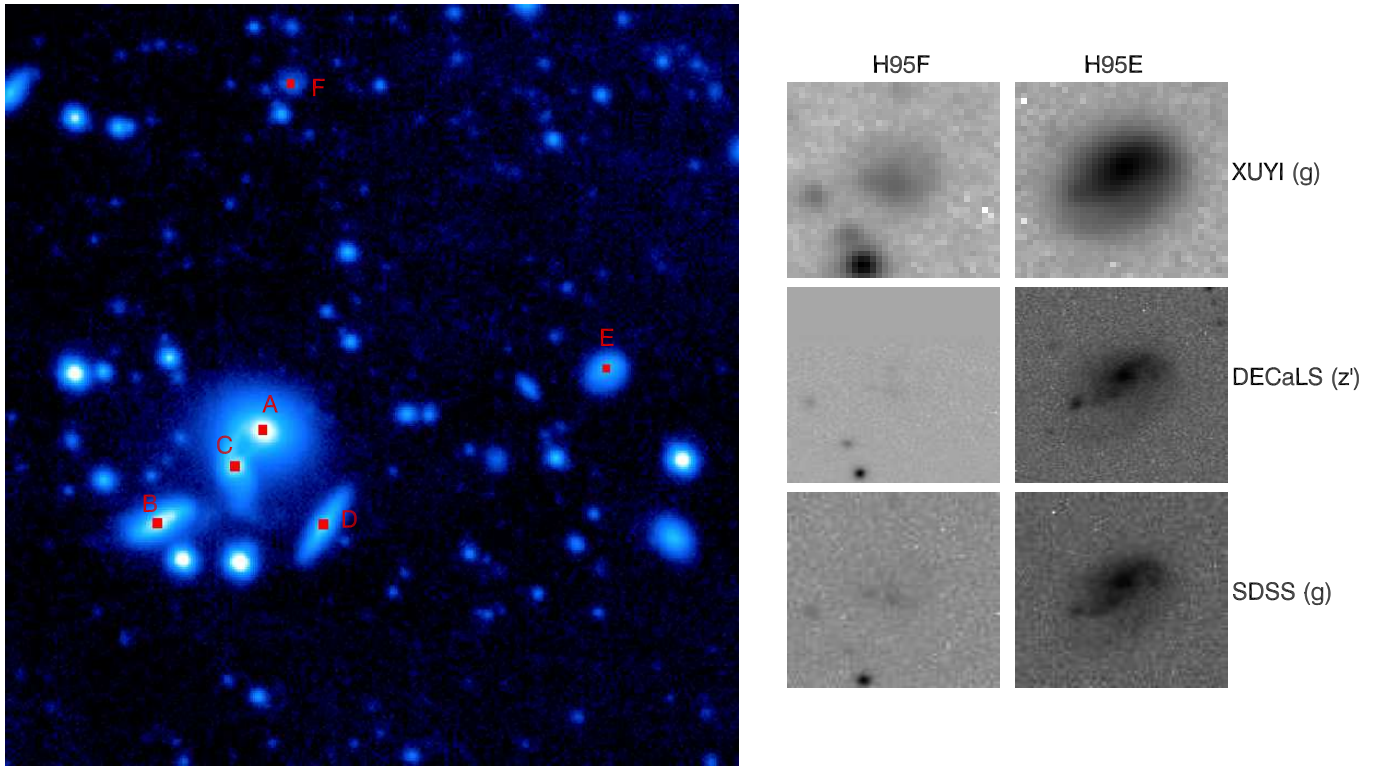} 
  \caption{Distribution of member galaxies for HCG\,95. Left: members of HCG\,95; the region is 6$\farcm$7 $\times$ 6$\farcm$7, and the top of red point is H95F, the UDG with H\,I distribution. Right: H95F and H95E are detected by the Xuyi telescope, DECaLS, and SDSS; the size of the postage stamp images is 37$\arcsec$.}
  \label{fig:fig3}
 \end{center}
\end{figure*}

Considering the uncertainties due to the variation of PSF FWHM in our observations, we examine our results by measuring the central surface brightness of UDGs using GALFIT \citep{peng2002}. The averaged PSF in the central region of our deep $g$-band science image is used. The central surface brightness is then extracted from the model S{\'e}rsic profile \citep{Graham2005,Zhong2008}. Finally, we obtain the $g$-band central surface brightness for 89 of the 105 UDGs, as shown in Figure~\ref{fig:fig2}.

The HCG\,95 field is covered by the Dark Energy Camera Legacy Survey (DECaLS) \citep{Blum2016}, which provides seeing-limited $z\arcmin$-band images obtained with the 4\,m Blanco telescope. Of the 89 UDG candidates, 84 have $z\arcmin$-band images from Data Release 3 (DR3)\footnote{\url{http://legacysurvey.org/dr3/description/}}. We measured the half-light radius from their $z\arcmin$-band images and list the results in Table~\ref{tab:taba1}, confirming that they have very LSB. Four examples of UDG candidates are shown in the right panel of Figure~\ref{fig:fig1}. 

Finally, we identify 89 UDG candidates in the HCG\,95 field. Redshift information is required to determine whether they are associated with HCG\,95. As shown in Figure~\ref{fig:fig3}, HCG\,95 is a compact group ($\alpha(\rm J2000) = 23^{h} 19^{m} 31.73^{s} $ and  $\delta(\rm J2000) = +9\degr\,29\arcmin\,30\farcs7$, redshift $z=0.0396$) with four bright galaxies seen in the central region. Of the four galaxies, H95A is a giant elliptical galaxy. H95C appears to contain double nuclei, therefore it is considered to be a merger remnant of two disk galaxies \citep{Iglesias-Paramo1997,Iglesias-Paramo1998}. H95C has two obvious tidal tails and a bridge connecting it to H95A \citep{Rodrigue1995}. H95B is a foreground galaxy because its line-of-sight velocity significantly differs from the velocity of the other members \citep{Iglesias-Paramo1997,DaRocha2005}. H95D is an edge-on spiral galaxy \citep{Iglesias-Paramo1997}. 
Many observations have been made of HCG\,95, including H\,{\small I}, H{\small$\alpha$} and X-ray \citep{Ponman1996,Iglesias-Paramo1998,Iglesias-Paramo2001,Huchtmeier2000,DaRocha2005}. Interestingly, three structures can be found  in the HCG\,95 field, including the Pegasus~I cluster \citep{Chincarini1976,O'Neil1997}, the Pegasus~II cluster \citep{Richter1982}, the Abell\,2588 cluster \citep{Richter1982,Abell1989}. These overdensities provide various environments in which to search for UDGs.

\begin{figure*}
 \setlength{\abovecaptionskip}{5pt}
 \begin{center}
  \includegraphics[trim=6mm 10mm 0mm 5mm,height=0.75\textwidth]{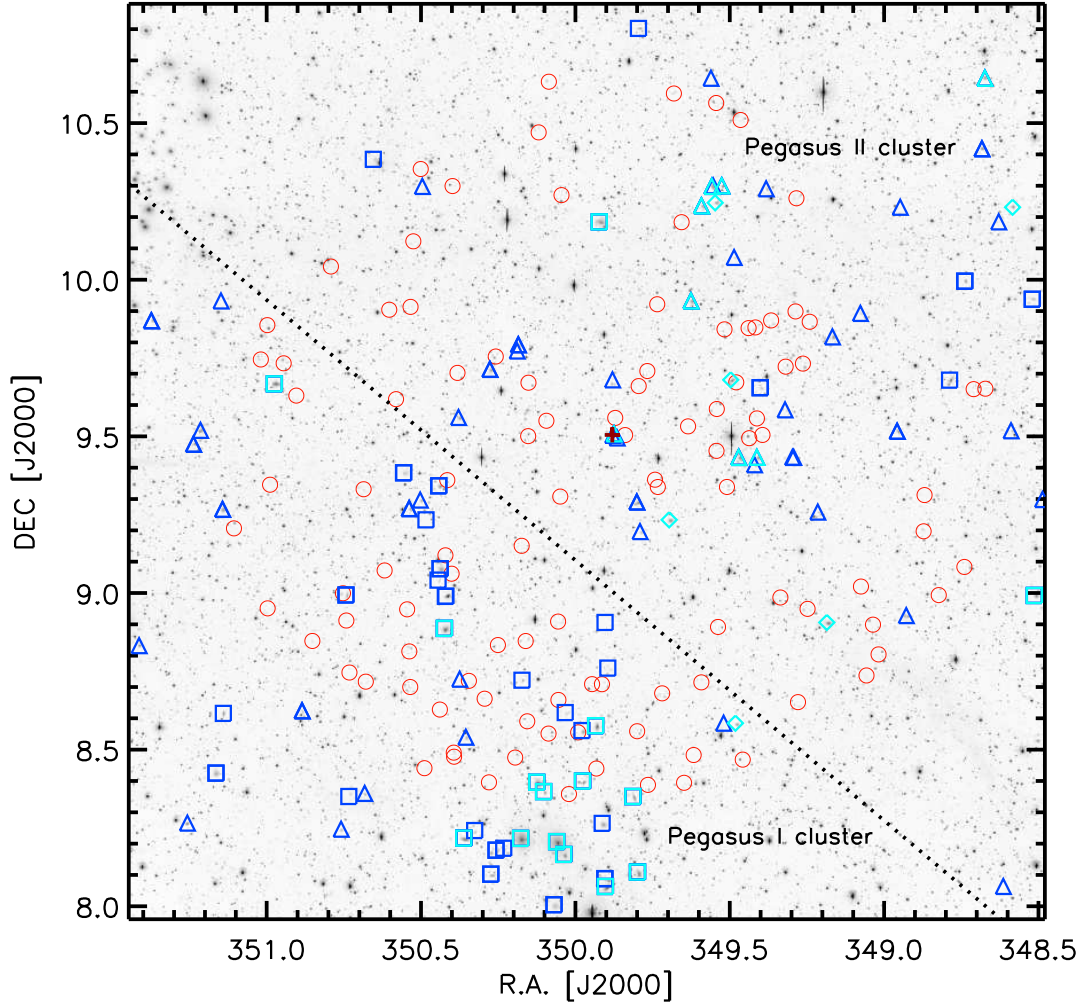}

  \caption{Spatial distribution of our 89 candidates of UDGs (red circles). The group HCG\,95 is marked by the plus at the center. Squares represent nearby galaxies with $0.009<z<0.016$. Triangles show nearby galaxies with $0.034<z<0.045$. These nearby galaxies are associated with the poor galaxy clusters Pegasus~I (squares) at $z=0.013$ and Pegasus~II (triangles) at $z=0.040$. Of these, 29 are included in the NGC catalog and are shown with cyan symbols. The dotted line is a crude guide to separate the two galaxy clusters.}
  \label{fig:fig4}
 \end{center}
\end{figure*}

\begin{figure}
 \setlength{\abovecaptionskip}{-5pt}
 \begin{center}
  \includegraphics[trim=5mm 0mm 0mm 10mm,clip,height=0.4\textwidth]{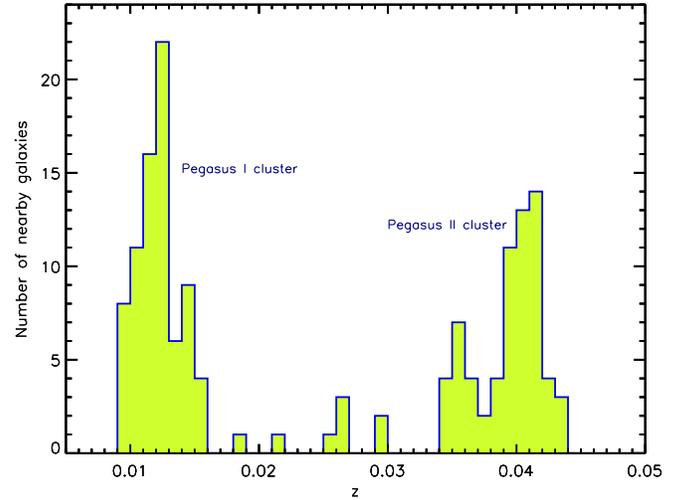}
  \caption{Histogram of nearby galaxies in the HCG\,95 field. These galaxies are divided into two distinct structures, we term them the Pegasus~I/Pegasus~II (Pegasus~I at $z$=0.013 and Pegasus~II at $z$=0.040) clusters.  }
  \label{fig:fig5}
 \end{center}
\end{figure}

In the 3$\degr$   $\times$   3$\degr$ field centered at HCG\,95, there are 96 nearby galaxies with known redshift from the literature, including 29 NGC galaxies and those from the Update Zwicky Catalogue (UZC) catalog \citep{Falco1999}, from the Flat Galaxy Catalog (FGC), the Uppsala Galaxy Catalog (UGC), from the SDSS, and from the Arecibo Legacy Fast ALFA (ALFALFA) \citep{Haynes2011,Teimoorinia2017}. The 29 NGC galaxies are listed in Table~\ref{tab:tab1}. The spatial distributions of these nearby galaxies are shown in Figure~\ref{fig:fig4}. 
We can see that these galaxies are located in two distinct redshift bins $0.009<z<0.016$ and $0.034<z<0.045$, which are associated with Pegasus~I ($z\sim 0.013$) and Pegasus~II ($z\sim 0.040$, similar to HCG\,95) \citep{Richter1982,Canizares1986}, as shown in Figure~\ref{fig:fig5}. Moreover, the two groups of galaxies are also globally separated from each other; this split is shown by the dotted line in Figure~\ref{fig:fig4}. We thus conclude that HCG\,95 is a part of the poor galaxy cluster Pegasus~II and our UDG candidates are most likely associated with either Pegasus~I at $z\sim0.013$ or Pegasus~II at $z\sim 0.040$. 
Indeed,  from Figure~\ref{fig:fig4}  one can see that the UDG candidates (red open circles) are also clustered with the two structures, supporting that the UDG candidates reside in the two poor galaxy clusters.

We note that our selection criteria for UDGs are based on the distance of HCG\,95 (i.e., the Pegasus~II cluster). However, Pegasus~I locates is located at a distance closer than Pegasus~II. Accordingly, the physical sizes of UDG candidates associated with Pegasus~I would be smaller. There are 44/45 UDG candidates located below/above the dotted line in Figure~\ref{fig:fig4},  likely associated with  Pegasus~I/Pegasus~II.  When the distance of Pegasus~I ($z=0.013$, $D=$55.2\,Mpc, $m-M=33.51$\,mag) is adopted to calculate the size and luminosity, 12 of the 44 UDG candidates still satisfy the UDG selection criterion $r_{\rm e}$\,$>$\,1.5\,kpc. The remaining 32 would be very LSB galaxies with sizes smaller than the size of a typical UDG. The 45 candidates above the dotted line would be UDGs if they are located close to Pegasus~II.  The sum of two parts  would mean that 57 UDGs are located in the HCG\,95 field.  We caution that member galaxies of the two clusters are widely spread over the field and information on the redshift is necessary to decide the real membership and physical size for a given UDG candidate. Still, the estimate of UDG numbers should be reasonable  in a statistic manner. For simplification, we assume that all 44 UDG candidates in the Pegasus~I region are at a distance of $z=0.013$ and all 45 UDG candidates in the Pegasus~II region at a distance of $z=0.040$.  Table~\ref{tab:taba1} lists the measured properties of our sample of 89 targets and denotes them to be either UDGs or very LSB galaxies.  In short, we conclude that roughly $\sim 50-60$ galaxies  are identified to be UDGs  associated with the two poor galaxy clusters in the HCG\,95 field.

\startlongtable
\begin{deluxetable}{ccccc}
\centering 
\tabletypesize{\scriptsize}
\tablecaption{Properties of the 29~NGC galaxies  in the HCG\,95 field collected  from the NASA Extragalactic Database (NED)\label{tab:tab1}. }
\tablewidth{0pt}
\tablehead{
     \colhead{Name}   & \colhead{R.A.}  & \colhead{Decl.}  & \colhead{$z$}  & \colhead{$m-M$}    \\
     \colhead{(\rm NGC/IC)}  &\colhead{(J2000.0)} & \colhead{(J2000.0)} &   & \colhead{(mag)}    }
     
      \startdata
      7612   & \colhead{23:19:44.2}  & \colhead{$+$08:34:35} &\colhead{0.0107}  & \colhead{32.97}     \\
      7611   & \colhead{23:19:36.7}  & \colhead{$+$08:03:48} &\colhead{0.0108}  & \colhead{33.00}     \\
      7634   & \colhead{23:21:41.9}  & \colhead{$+$08:53:14} &\colhead{0.0108}  & \colhead{32.98}     \\
      7626   & \colhead{23:20:42.6}  & \colhead{$+$08:13:01} &\colhead{0.0114}  & \colhead{33.11}     \\
      7608   & \colhead{23:19:15.3}  & \colhead{$+$08:21:02} &\colhead{0.0117}  & \colhead{33.18}     \\
      7610\tablenotemark{a}   & \colhead{23:19:41.5}  & \colhead{$+$10:11:05} &\colhead{0.0119}  & \colhead{33.22}     \\
      7648\tablenotemark{b}   & \colhead{23:23:54.1}  & \colhead{$+$09:40:04} &\colhead{0.0119}  & \colhead{33.22}     \\
      7631   & \colhead{23:21:26.8}  & \colhead{$+$08:13:03} &\colhead{0.0125}  & \colhead{33.35}     \\
      7623   & \colhead{23:20:30.1}  & \colhead{$+$08:23:46} &\colhead{0.0125}  & \colhead{33.34}     \\
      7619   & \colhead{23:20:14.6}  & \colhead{$+$08:12:23} &\colhead{0.0125}  & \colhead{33.35}     \\
      7621   & \colhead{23:20:24.7}  & \colhead{$+$08:21:57} &\colhead{0.0128}  & \colhead{33.40}     \\
      7617   & \colhead{23:20:09.1}  & \colhead{$+$08:09:56} &\colhead{0.0139}  & \colhead{33.60}     \\
      5309   & \colhead{23:19:11.8}  & \colhead{$+$08:06:33} &\colhead{0.0140}  & \colhead{33.62}     \\
      7615   & \colhead{23:19:54.5}  & \colhead{$+$08:23:58} &\colhead{0.0149}  & \colhead{33.77}     \\
      7529   & \colhead{23:14:03.2}  & \colhead{$+$08:59:32} &\colhead{0.0151}  & \colhead{33.81}     \\      
      7569   & \colhead{23:16:44.5}  & \colhead{$+$08:54:22} &\colhead{0.0217}  & \colhead{34.66}     \\      
      5306   & \colhead{23:18:11.3}  & \colhead{$+$10:14:44} &\colhead{0.0253}  & \colhead{35.02}     \\
      7586   & \colhead{23:17:55.6}  & \colhead{$+$08:35:04} &\colhead{0.0267}  & \colhead{35.14}     \\
      7601   & \colhead{23:18:47.1}  & \colhead{$+$09:14:01} &\colhead{0.0268}  & \colhead{35.15}     \\
      7528   & \colhead{23:14:20.1}  & \colhead{$+$10:13:53} &\colhead{0.0292}  & \colhead{35.35}     \\
      7587   & \colhead{23:17:59.3}  & \colhead{$+$09:40:49} &\colhead{0.0292}  & \colhead{35.35}     \\     
      5305   & \colhead{23:18:06.3}  & \colhead{$+$10:17:59} &\colhead{0.0350}  & \colhead{35.77}     \\
7594\tablenotemark{c}   & \colhead{23:18:14.0}  & \colhead{+10:17:52} &\colhead{0.0362}  & \colhead{35.84}     \\
      7542   & \colhead{23:14:41.7}  & \colhead{$+$10:38:35} &\colhead{0.0395}  & \colhead{36.05}     \\
      7609   & \colhead{23:19:29.9}  & \colhead{$+$09:30:29} &\colhead{0.0396}  & \colhead{36.05}     \\
      7595   & \colhead{23:18:30.2}  & \colhead{$+$09:55:56} &\colhead{0.0410}  & \colhead{36.13}     \\
      7579   & \colhead{23:17:38.9}  & \colhead{$+$09:25:59} &\colhead{0.0411}  & \colhead{36.14}     \\
      5307   & \colhead{23:18:22.0}  & \colhead{$+$10:14:09} &\colhead{0.0413}  & \colhead{36.17}     \\
      7584   & \colhead{23:17:53.1}  & \colhead{$+$09:25:58} &\colhead{0.0431}  & \colhead{36.25}     
\enddata
\tablenotetext{a}{NGC~7610 and NGC~7616 refer to the same object;}
\tablenotetext{b}{NGC~7648 = IC~1486;}
\tablenotetext{c}{NGC~7594 = IC~1478.}
\end{deluxetable}

\begin{figure}
 \setlength{\abovecaptionskip}{-15pt}
\begin{center}
\includegraphics[trim=25mm 0mm 20mm 5mm,clip,height=0.55\textwidth]{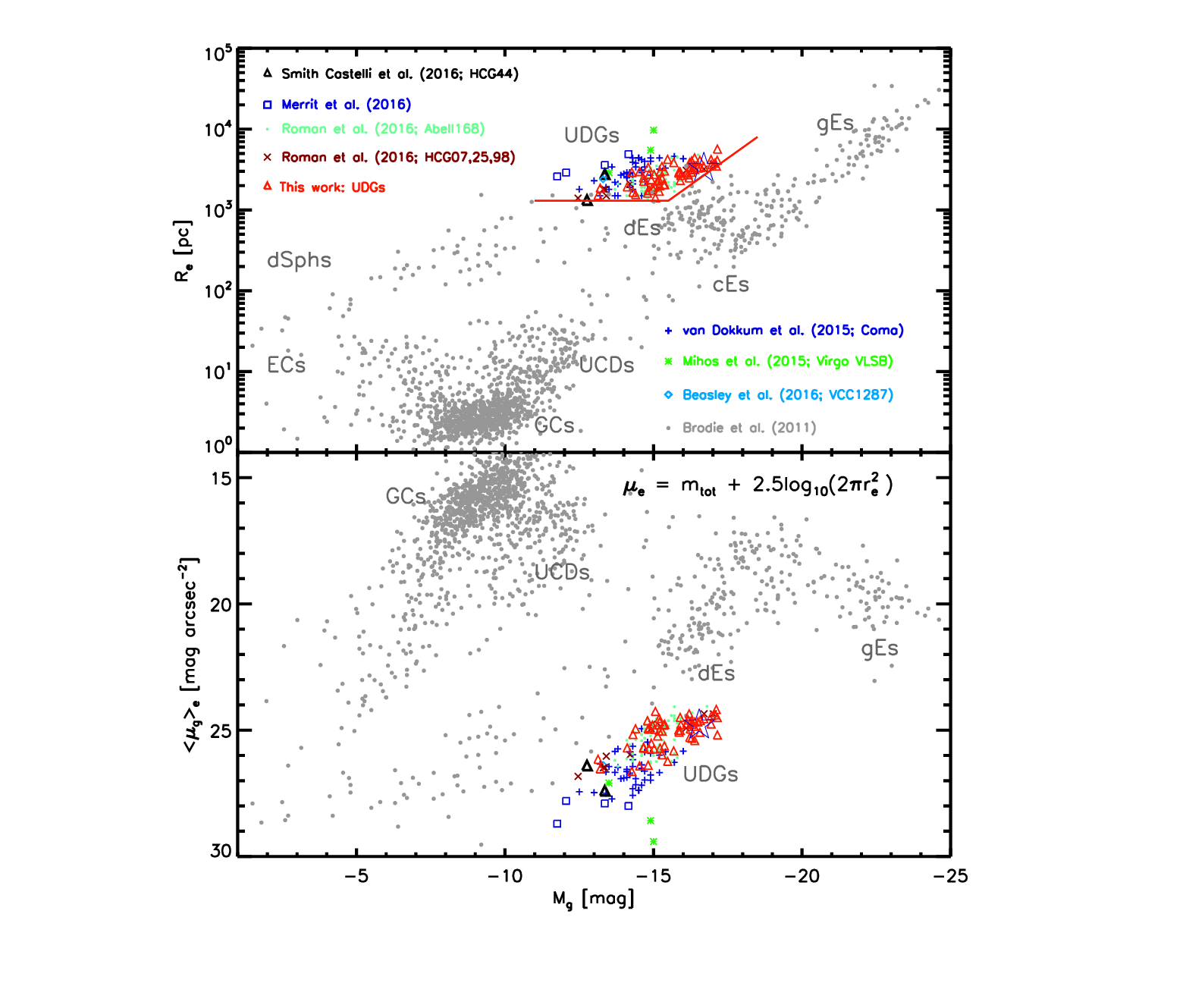}
\caption{Comparison of UDGs with other galaxy populations in diagrams of absolute magnitude in $g$ vs. effective radius and average surface brightness. The gray dots are quiescent galaxies from \cite{Brodie2011}; blue pluses represent 47 UDGs from the Coma cluster \citep{van Dokkum2015a}; green tail symbols are the extremely LSB galaxies from \cite{Mihos2015}; the light blue diamond refers to the single UDG from \cite{Beasley2016}; black triangles show 2 UDGs found in the HCG 44 field \citep{Smith2016}; blue squares represent the 4 UDGs in the M101 field \citep{Merritt2016};  light green dots are 80 UDGs from Abell\,168 \citep{Roman2017}; brown crosses are 11 UDGs from HCG~07, 25, and 98 \citep{RomanTrujillo2016}; and red triangles denote UDGs from this work.}
\label{fig:fig6}
\end{center}
\end{figure}

\begin{figure}
\begin{center}
\includegraphics[trim=15mm 1mm 0mm 5mm,clip,height=0.35\textwidth]{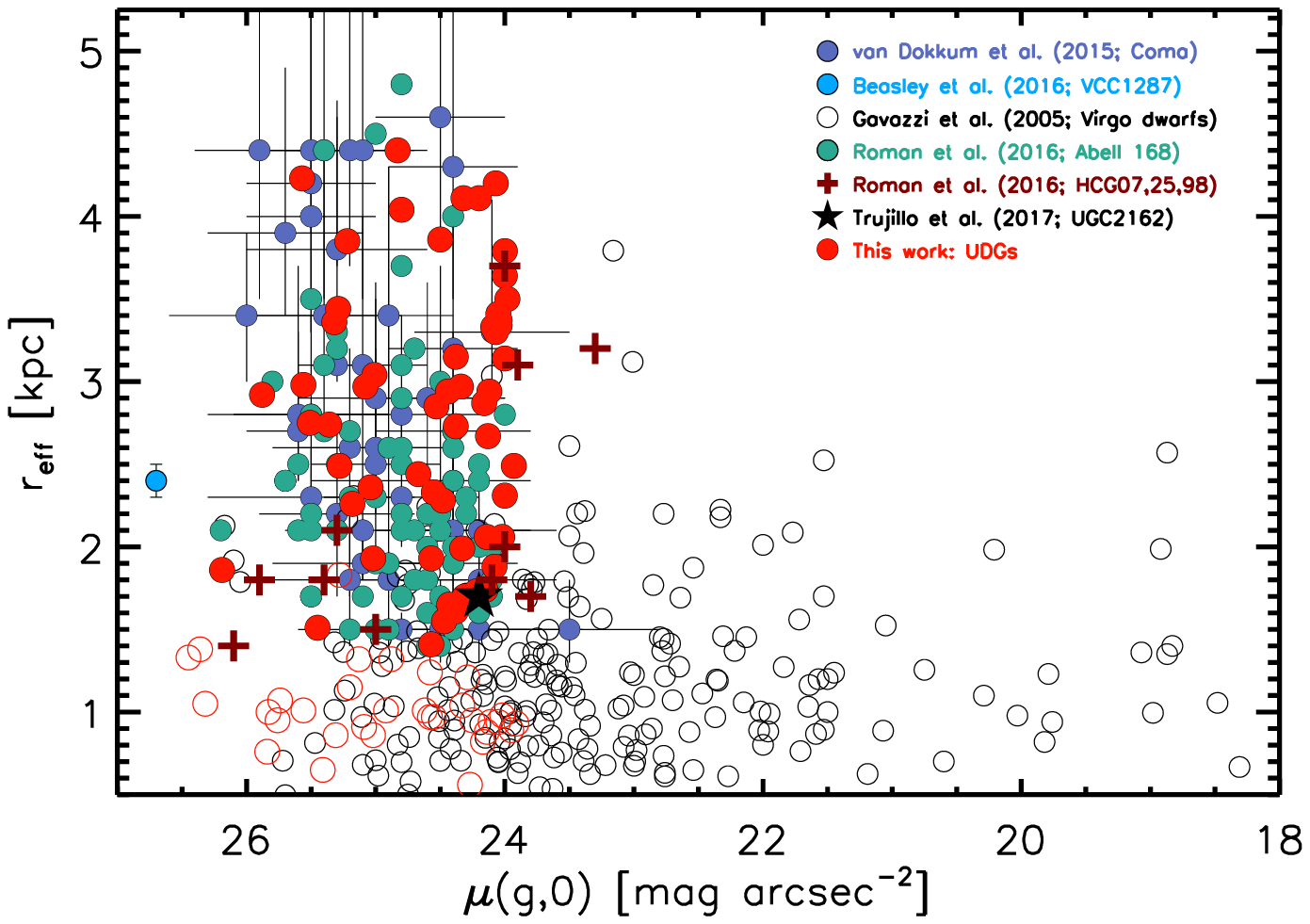}
\caption{Relationship between half-light radius and $g$-band central surface brightness. The black open circles represent Virgo dwarfs \citep{Gavazzi2005}; the blue solid circles are UDGs from Coma cluster \citep{van Dokkum2015a}; indigo solid circles are UDGs from the Abell\,168 cluster \citep{Roman2017}; the light blue solid circle refers to the UDG in the Virgo cluster from \cite{Beasley2016}; and brown pluses represent 11 UDGs from HCG~07, HCG~25, and HCG~98 \citep{RomanTrujillo2016}. Red solid circles mark our UDGs and  red open circles represent the very LSB galaxies (possible UDGs) in this work.}
\label{fig:fig7}
\end{center}
\end{figure}

\begin{figure}
\setlength{\abovecaptionskip}{-2pt}
\begin{center}
\includegraphics[trim=1mm 1mm 0mm 5mm,clip,height=0.35\textwidth]{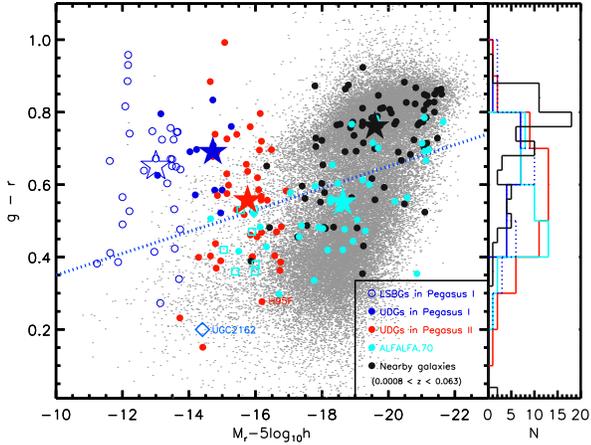}
\caption{Diagram of color $g-r$ vs. absolute magnitude in $r$ for our UDGs and nearby galaxies. The gray points represent the galaxies with $0.001<z<0.06$ from the SDSS. The blue dotted line is adopted from \cite{Blanton2006}  to split the SDSS galaxies into red sequence and blue cloud.  Our sample of 89 UDG candidates are divided into subsamples, including 32 extremely LSB galaxies in the Pegasus~I region (blue open circles), 12 UDGs at the distance of Pegasus~I (blue solid circle), and 45 UDGs at the distance of Pegasus~II (red solid circles).  Galaxies with rich H\,{\small I} gas from the ALFALFA survey are shown with cyan solid circles. The cyan squares are the UDGs with strong H\,{\small I} emission from the ALFALFA catalog \citep{Leisman2017}.  The confirmed member galaxies  of two poor galaxy clusters in the HCG\,95 field are also shown (black solid circles). The median of $g-r$ for each subsample of UDGs is marked with large star symbols. The light blue square is a blue UDG from \cite{Trujillo2017}. The right panel presents the number distribution of color from  these subsample.}
\label{fig:fig8}
\end{center}
\end{figure}

\begin{figure}
 \setlength{\abovecaptionskip}{5pt}
 \centering
 \includegraphics[trim=2mm 2mm 0mm 5mm,clip,height=0.4\textwidth]{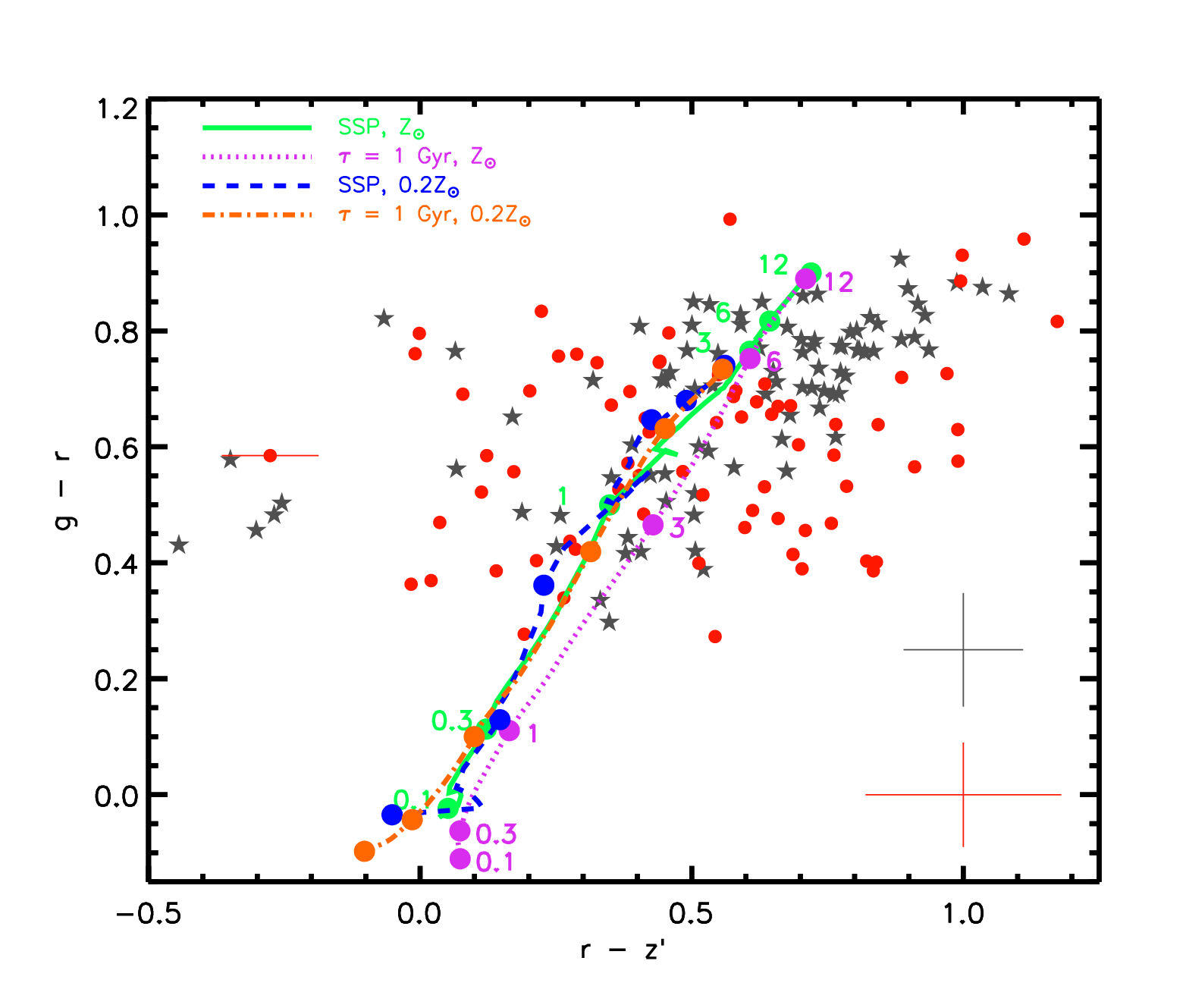}
\caption{Color-color diagram of UDGs and spectroscopically identified nearby galaxies in the HCG\,95 field. The gray stars represent the nearby galaxies with redshifts $0.009<z<0.05$. The red solid circles show the 89 UDG candidates. BC03 model color tracks are shown for the single stellar population. The green solid line and the magenta dotted line denote the evolution tracks for SSP at one solar metallicity. The blue dotted line and the orange dash-dotted line show the evolution tracks for SSP at metallicity 0.2$Z_{\odot}$. The magenta dotted line and orange dash-dotted line give different metallicty $\tau$ models with $\tau$ = 1 Gyr. The gray and red error bars at the bottom right denote the $0.5\sigma$ dispersion of the galaxies.}
\label{fig:fig9}
\end{figure}

We examine the properties of our sample of UDGs and very LSB galaxies in comparison with other samples of UDGs in the literature. Figure~\ref{fig:fig6} shows the absolute magnitude $M_{g}$ in relation to the average surface brightness $\mu_{\rm e}$  and effective radius $r_{\rm e}$.   Figure~\ref{fig:fig7} presents the central surface brightness $\mu{(g,0)}$ in relation to the effective radius $r_{\rm e}$. 
From Table~\ref{tab:taba1} we can see that the 32 very LSB galaxies in our sample are characterized by $\mu(g,0)$\,=\,24.00\,$-$\,26.45\,mag\,arcsec$^{-2}$, typical $<r_{\rm e}>$\,$\sim$\,0.98\,kpc, typical $<M_{g}>$\,$\sim$\,$-$13.24\,mag, color $<g-r>\,\sim\,0.65$, and axis ratio $<b/a>\,\sim\,0.75$; while the 12 UDGs have $\mu(g,0)$\,=\,24.08\,$-$\,26.19\,mag\,arcsec$^{-2}$, $<r_{\rm e}>$\,$\sim$\,1.93\,kpc, $<M_{g}>$\,$\sim$\,$-$14.66\,mag, $<g-r>\,\sim\,0.69$ and $<b/a>\,\sim\,0.71$;  and the 45 UDGs in Pegasus~II have $\mu(g,0)$\,=\,24.00\,$-$\,25.88\,mag\,arcsec$^{-2}$, $<r_{\rm e}>$\,$\sim$\,2.98\,kpc, $<M_{g}>$\,$\sim$\,$-$15.90\,mag, $<g-r>\,\sim\,0.56$ and $<b/a>\,\sim\,0.67$. 
The comparisons in Figures~\ref{fig:fig6} and \ref{fig:fig7} confirm that the selected UDGs in the HCG\,95 field are distinct from normal galaxies in terms of these parameters and comparable to those found in previous studies \citep[e.g.,][]{vanderBurg2016}.  The consistence indicates that UDGs can be found in different environments. 

Figure~\ref{fig:fig8} presents the $r$-band absolute magnitude versus $g-r$ for our UDGs in comparison with other galaxy populations, including nearby galaxies from the SDSS and H\,{\small I} gas-rich galaxies from the ALFALFA survey. Our sample UDGs are spread over a wide range in $g-r$, similar to the spread of member galaxies of the poor galaxy clusters Pegasus~I and Pegasus~II. A sample of nearby mass-selected galaxies with $0.001<z<0.06$ are also presented for comparison. It is clear from Figure~\ref{fig:fig8} that our sample contains  $>$20 UDGs with colors similar to the colors of star-forming galaxies, although most UDGs are as red as quiescent dwarf galaxies.  The UDGs in the Pegasus~II cluster appear to be slightly bluer in $g-r$ than the UDGs in the Pegasus~I cluster. The diversity of color indicates that UDGs may be formed through different star formation histories, which are to some extent coupled with the environments.  

Figure~\ref{fig:fig9} shows the $r-z\arcmin$ versus $g-r$ diagram for 89 UDG candidates and 96 spectroscopically identified nearby galaxies in the HCG\,95 field. We can see that the 89 UDG candidates tend to be fainter and bluer than the nearby galaxies (as shown in Figure~\ref{fig:fig8}), although the dispersion in two colors is rather large. We also show the color tracks of galaxy models of different ages from BC03 \citep{Bruzual2003}. The single stellar population (SSP) and an e-folding ($\tau$=1\,Gyr) evolutionary history are adopted to generate the templates. The metallicity is set to be solar and 1/5 solar for comparison. The typical ages of 0.1, 0.3, 1, 3, 6, 12\,Gyr are marked with solid circles. We note that the $z\arcmin$-band photometry has large uncertainties for the faint UDG candidates and the dispersion in color $r-z\arcmin$ is largely driven by the measurement errors. We note that the model templates are only for stellar emission. The presence of strong emission lines in the $r$ band (i.e. H$\alpha$\,+\,[NII]) would apparently make galaxies appear bluer in $r-z\arcmin$ and redder in $g-r$. It is obvious that a large fraction of UDG candidates have young stellar populations ($<\sim$3\,Gyr), and nearly one-third are as blue as blue-cloud galaxies.                                                            

We derived the stellar mass $M_{\star}$ from luminosity and color using the mass-to-light ratio versus $g-r$ relation \citep{Bell2003,Taylor2011}. The typical stellar mass  is $M_{\star} \sim 4.32 \times 10^{7}\,M_{\odot}, 1.58 \times 10^{8}\,M_{\odot}$, and 3.72 $\times 10^{8}\,M_{\odot}$ for 32 very LSB galaxies, 12 UDGs in Pegasus~I, and 45 UDGs in Pegasus~II, respectively. The stellar masses of UDGs are consistent with previous studies \citep{van Dokkum2015a,Roman2017,RomanTrujillo2016}. 
Combined together, our results  suggest that UDGs not only exist in galaxy clusters such as Coma, Virgo, Fornax, and Abell \citep{Mihos2015,Munoz2015,Koda2015,van Dokkum2015a,van Dokkum2015b,Martinez-Delgado2016,Janssens2017}, but also live  in lower density environments.

Observations of neutral hydrogen gas are available for the galaxy group HCG\,95 \citep{Huchtmeier2000}. Two member dwarf galaxies  H95E and H95F about 3$\farcm$5 offset from the center of HCG\,95 (about 164.6\,kpc at the distance of HCG\,95), as shown in Figure~\ref{fig:fig3}, have been detected to have strong H\,{\small I} emission with the Very Large Array (VLA). The line-of-sight velocities of the two galaxies are $11830 \pm 5$ km\,s$^{-1}$ and  $11730 \pm 5$ km\,s$^{-1}$, respectively. From the VLA observations, we derive the H\,{\small I} gas mass for H95E and H95F using the formula ($M_{\rm HI}/M_{\odot}) = 2.36\times 10^{5} (f/$Jy\,km\,s$^{-1})(D$/Mpc)$^{2}$ \citep{Levy2007}, where $f$ is the 21\,cm line flux (in Jy\,km\,s$^{-1}$) and $D$ is the distance of HCG\,95. The derived H\,{\small I} mass is 2.02  and 1.08 $\times 10^{9}\,M_{\odot}$ for H95E and H95F, respectively. 

Interestingly, H95F is one of our sample UDGs. The properties of H95F are given in Table~\ref{tab:tab2}.  In addition, H95F is a blue dwarf galaxy with $ g-r \sim 0.28$ and $M_{\star}\sim 1.82\times10^{8}$\,$M_{\odot}$. We further address the connections between UDGs and gas-rich LSB galaxies.  \citet{Impey1996} have compiled a sample of 693 LSB galaxies at $z \la 0.1$. We find that 16 extended LSB galaxies in this sample meet the selection criteria of UDGs, and  15 of them have an H\,{\small I} detection. The physical parameters of the 16 UDGs along with H95F are given in Table~\ref{tab:tab2}. They have H\,{\small I} gas masses from $5.13\times 10^{7}$\,$M_{\odot}$ to 3.39$\times 10^{9}$\,$M_{\odot}$ and stellar masses over the range of 10$^{6} - 10^{8}$\,$M_{\odot}$ derived from color and luminosity \citep{Fukugita1996,Bell2003,Taylor2011}. Recently, \citet{Trujillo2017} have reported one nearest UDG (UGC~2162) with strong H\,{\small I} emission; the properties of this UDG are also listed in Table~\ref{tab:tab2}. \cite{Leisman2017} presented 115 UDGs with H\,{\small I} emission from the ALFALFA survey. We include 7 of the 115 UDGs in Table~\ref{tab:tab2} for a comparison. The overlap of UDGs with some extremely LSB galaxies detected by H\,{\small I} surveys in size and central surface brightness reveals an intimate connection between the two populations, providing key insights into the formation of UDGs in the local universe \citep{Bellazzini2017}. 

\begin{deluxetable*}{cchlDllllc}
\tabletypesize{\scriptsize}
\tablenum{2}
\tablecaption{Comparison of H95F with the extremely LSB galaxies (Can Be Seen as UDGs) with H\,{\small I} gas detection from \citet{Impey1996}. \label{tab:tab2}}
\tablewidth{0pt}
\tablehead{
     \colhead{Name}   & \colhead{R.A.}  & \colhead{Decl.}  & \colhead{$\mu(B,0)$}  & \colhead{$m_{B}$} & \colhead{$M_{B}$} & \colhead{$\rm Velocity$} & \colhead{$r_{\rm e}$} & \colhead{$r_{\rm e}$}  & \colhead{$\log(M_{\rm H\,I})$}  \\
          &\colhead{(J2000.0)} & \colhead{(J2000.0)} &\colhead{(mag\,arcsec$^{-2}$)}  & \colhead{(mag)} & \colhead{(mag)}  & \colhead{(km\,s$^{-1}$)} & \colhead{($\arcsec$)}  & \colhead{(kpc)} & \colhead{($M_{\odot}$)}  }

\startdata
      H95F$\tablenotemark{a}$        & \colhead{23:19:29.0} & \colhead{$+$9:33:31}    &\colhead{24.1}   & \colhead{19.4} & \colhead{$-$16.7}  & \colhead{11730} & \colhead{4.35} & \colhead{3.41} & \colhead{9.03}    \\ 
      UGC~2162$\tablenotemark{b}$     & \colhead{2:40:23.1}  & \colhead{$+$01:13:45}   &\colhead{24.4}   & \colhead{16.1} & \colhead{$-$15.0}  & \colhead{1172} & \colhead{28} & \colhead{1.7} & \colhead{8.28}    \\
      322019$\tablenotemark{c}$      & \colhead{22:58:26.9} & \colhead{$+$01:50:58.9} &\colhead{24.6}  & \colhead{18.5} & \colhead{$-$15.8}  & \colhead{4819} & \colhead{11.9} & \colhead{3.9} & \colhead{8.81}    \\ 
      103796$\tablenotemark{c}$      & \colhead{00:20:39.6} & \colhead{$+$06:57:56.8} &\colhead{24.2}  & \colhead{18.3} & \colhead{$-$16.2}  & \colhead{5647} & \colhead{9.2} & \colhead{3.5} & \colhead{8.86}    \\ 
      113790$\tablenotemark{c}$      & \colhead{01:13:02.1} & \colhead{$+$27:38:12.8} &\colhead{24.3}  & \colhead{18.8} & \colhead{$-$15.4}  & \colhead{4952} & \colhead{7.1} & \colhead{2.4} & \colhead{8.57}    \\ 
      114905$\tablenotemark{c}$      & \colhead{01:25:18.5} & \colhead{$+$07:21:37}   &\colhead{24.9}  & \colhead{18.2} & \colhead{$-$16.2}  & \colhead{5435} & \colhead{12.8} & \colhead{4.7} & \colhead{9.11}    \\ 
      114943$\tablenotemark{c}$      & \colhead{01:47:06.6} & \colhead{$+$07:19:51.9} &\colhead{24.5}  & \colhead{18.9} & \colhead{$-$16.4}  & \colhead{8416} & \colhead{8.0} & \colhead{4.5} & \colhead{9.10}    \\ 
      113949$\tablenotemark{c}$      & \colhead{01:49:38.6} & \colhead{$+$30:40:50.8} &\colhead{24.3}  & \colhead{19.2} & \colhead{$-$15.8}  & \colhead{7380} & \colhead{5.0} & \colhead{2.5} & \colhead{9.03}    \\ 
      122966$\tablenotemark{c}$      & \colhead{02:09:29.0} & \colhead{$+$31:51:10}   &\colhead{25.4}  & \colhead{18.4} & \colhead{$-$16.4}  & \colhead{6518} & \colhead{11.6} & \colhead{5.1} & \colhead{9.00}    \\ 
      \hline
      0139+0240   & \colhead{1:39:58.9}   & \colhead{$+$2:40:40} &\colhead{24.4}  & \colhead{16.9} & \colhead{$-$14.3}  & \colhead{1765} & \colhead{26.0} & \colhead{3.15} & \colhead{8.60}    \\
      0221+0034   & \colhead{2:21:49.8}   & \colhead{$+$0:34:41} &\colhead{24.7}  & \colhead{17.3} & \colhead{$-$17.5}  & \colhead{8996} & \colhead{18.9} & \colhead{11.35}& \colhead{9.25}    \\
      0225$-$0049   & \colhead{2:25:45.6}   & \colhead{$-$0:49:50} &\colhead{25.0}  & \colhead{18.0} & \colhead{$-$12.8}  & \colhead{1464} & \colhead{13.2} & \colhead{1.33} & \colhead{7.71}    \\
      0227+0040   & \colhead{2:27:01.1}   & \colhead{$+$0:40:56} &\colhead{24.8}  & \colhead{18.1} & \colhead{$-$16.6}  & \colhead{8503} & \colhead{13.7} & \colhead{7.79} & \colhead{9.53}    \\
      0249+0146   & \colhead{2:49:45.0}   & \colhead{$+$1:46:16} &\colhead{24.5}  & \colhead{16.5} & \colhead{$-$16.7}  & \colhead{4293} & \colhead{21.2} & \colhead{6.19} & \colhead{8.87}    \\
      0319+0015   & \colhead{3:19:14.2}   & \colhead{$+$0:15:00} &\colhead{24.4}  & \colhead{18.4} & \colhead{$-$15.7}  & \colhead{6548} & \colhead{8.3}  & \colhead{3.66} & \colhead{9.24}    \\
      0955+0155   & \colhead{9:55:54.5}   & \colhead{$+$1:55:58} &\colhead{24.9}  & \colhead{17.2} & \colhead{$-$14.1}  & \colhead{1815} & \colhead{18.3} & \colhead{2.28} & \colhead{8.54}    \\
      1101+0211   & \colhead{11:01:45.4}  & \colhead{$+$2:11:24} &\colhead{24.1}  & \colhead{18.2} & \colhead{$-$16.2}  & \colhead{7578} & \colhead{10.9} & \colhead{5.54} & \colhead{9.17}    \\
      1110$-$0017   & \colhead{11:10:14.1}  & \colhead{$-$0:17:47} &\colhead{24.8}  & \colhead{17.8} & \colhead{$-$16.9}  & \colhead{8472} & \colhead{11.6} & \colhead{6.57} & \colhead{9.37}    \\
      1154+0203   & \colhead{11:54:48.6}  & \colhead{$+$2 03:35} &\colhead{24.5}  & \colhead{17.7} & \colhead{$-$13.8}  & \colhead{1980} & \colhead{12.1} & \colhead{1.64} & \colhead{8.15}    \\
      1228+0116   & \colhead{12:28:43.1}  & \colhead{$+$1:16:13} &\colhead{25.1}  & \colhead{18.3} & \colhead{$-$13.5}  & \colhead{2289} & \colhead{11.1} & \colhead{1.74} & \colhead{8.20}    \\
      1230$-$0015   & \colhead{12:30:34.2}  & \colhead{$-$0:15:28} &\colhead{24.1}  & \colhead{16.5} & \colhead{$-$16.1}  & \colhead{3279} & \colhead{25.0} & \colhead{5.60} & \colhead{8.66}    \\
      1249+0233   & \colhead{12:49:49.9}  & \colhead{$+$2:33:34} &\colhead{25.0}  & \colhead{19.1} & \colhead{$-$15.1}  & \colhead{6971} & \colhead{10.4} & \colhead{4.84} & \colhead{8.91}    \\
      1350+0230   & \colhead{13:50:59.7}  & \colhead{$+$2:30:20} &\colhead{24.8}  & \colhead{18.5} & \colhead{$-$14.8}  & \colhead{4507} & \colhead{11.4} & \colhead{3.49} & \colhead{8.62}    \\
      1431+0142   & \colhead{14:31:20.5}  & \colhead{$+$1:42:25} &\colhead{25.8}  & \colhead{17.4} & \colhead{$-$14.0}  & \colhead{1829} & \colhead{24.2} & \colhead{3.04} & \colhead{8.30}    \\
      1307+0112   & \colhead{13:07:31.2}  & \colhead{$+$1:12:53} &\colhead{24.2}  & \colhead{18.2} & \colhead{$-$15.6}  & \colhead{5842} & \colhead{9.7}  & \colhead{3.83} & \colhead{\nodata}    \\
\hline
\enddata
\tablenotetext{a}{H95F is shown in this work.}
\tablenotetext{b}{Measured in the $g$ band, UGC~2162 from \cite{Trujillo2017}.}
\tablenotetext{c}{Measured in the $g$ band from \cite{Leisman2017}.}

\end{deluxetable*}

Motivated by the finding that H95F is an H\,{\small I} gas-rich galaxy, we cross-correlated our sample of 89 UDG candidates with the H\,{\small I} galaxy catalog from the ALFALFA survey \citep{Giovanelli2005,Haynes2011,Teimoorinia2017}. Although 49 nearby galaxies  with $0.009 < z < 0.058$ in the HCG\,95 field are found to have an H\,{\small I} detection by ALFALFA, none of our sample UDGs is detected in ALFALFA.  Deeper H\,{\small I} surveys will probably detect more UDGs with  H\,{\small I} gas.

We compare the abundances of UDGs of three galaxy structures in the HCG\,95 field with those of other groups and clusters from the literature. Figure~\ref{fig:fig10} shows the abundance of UDGs as a function of halo mass adopted from \cite{Janssens2017} and \cite{vanderBurg2016}, following a relation of $ N \propto M^{0.93\pm0.16}_{200}$. It is clear that more massive halos host more numerous UDGs. We use the velocity dispersion to estimate the halo masses of the three structures in the HCG\,95 field. The halo masses of galaxy group HCG\,95 and the two poor galaxy clusters Pegasus~I and II are 2 $\times 10^{13}\,M_{\odot}$, 3 $\times 10^{14}\,M_{\odot}$, and 5 $\times 10^{14}\,M_{\odot}$, respectively \citep{Chincarini1976,Girardi1998,Barnes1999,Randall2009,Valtchanov1999,DaRocha2005}. The abundance of UDGs in HCG\,95 is consistent with the overall correlation. However, the clusters Pegasus~I cluster and Pegasus~II cluster are located below the correlation. We point out that our observations cover only part of the sky area of the two clusters, and the estimated abundance of UDGs is therefore only the lower limit for the abundance of UDGs. The red arrows in Figure~\ref{fig:fig10} denote the lower limits for the two poor clusters. We thus argue that the three galaxy structures in the HCG\,95 field follow the same correlation between the abundance of UDGs and halo mass as other galaxy groups and clusters.

\begin{figure}
\setlength{\abovecaptionskip}{-2pt}
\begin{center}
\includegraphics[trim=1mm 1mm 0mm 5mm,clip,height=0.35\textwidth]{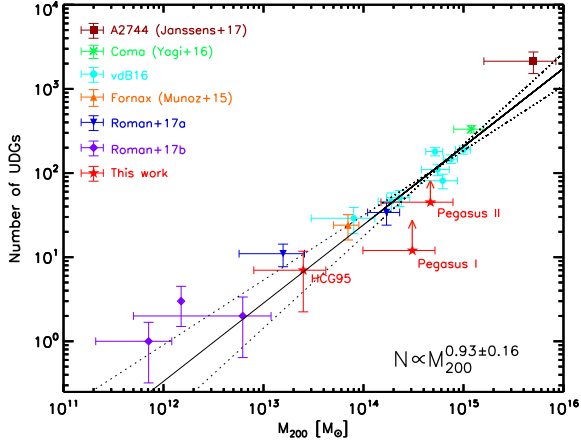}
\caption{Abundance of UDGs as a function of halo mass. The red solid stars show the abundance of UDGs from this work, while the other symbols are obtained from the other clusters and groups. The solid and dashed lines are the best-fit relation from \cite{Janssens2017}.}
\label{fig:fig10}
\end{center}
\end{figure}

\section{Discussion} \label{sec:discussion}

One uncertainty in our analysis is the distance for the selection of UDGs. We identified 89 UDG candidates over a sky area of 4.9 degree$^2$ assuming that they are located at the distance of HCG\,95, which resides in the galaxy cluster Pegasus~II \citep{Rood1994}. Given that another poor galaxy cluster, Pegasus~I, also exists in this field, some candidates may probably be linked with Pegasus~I at a distance closer than Pegasus~II. Then the actual physical size of these objects would be smaller. The extremely LSB of these candidates makes the measurement of their distance from spectroscopic observations very difficult even with current 10\,m class telescopes.  In terms of the spatial distribution of bright nearby galaxies with known redshift, the UDG candidates are crudely divided and assigned to the two poor clusters.  We estimated that roughly 50 $-$ 60 UDGs are present in the HCG\,95 field.  This confirms that UDGs exist in environments with a density that is lower than the density of rich galaxy clusters such as Coma.

The Pegasus~I cluster is embedded in the filamentary structures connecting the  Pisces-Perseus supercluster (PPS) and the local supercluster \citep{Richter1982,Levy2007}. The velocity dispersion  of the Pegasus~I cluster  is $\sigma \sim 236\pm33$\,km\,s$^{-1}$, which is one of the lowest velocities among galaxy clusters \citep{Noonan1981,Richter1982}.  We note that there are far fewer UDGs in the Pegasus~I region than in the Pegasus~II (HCG\,95) region. One reason for this is that the Pegasus~I cluster is closer, so that the area covered in our observations is smaller, although the angular diameter of the Pegasus~I cluster is much larger than the angle of the Pegasus~II cluster \citep{Chincarini1976}.  However, we cannot deny that the intrinsic abundance of UDGs in the two clusters may differ from each other given their different halo structures. 
The Pegasus~I and Pegasus~II clusters both provide a low-density environment. Galaxies in such low-density environments are systematically younger than those in dense environments and their formation timescale is  longer \citep{Thomas05}.  The newly formed disk galaxies of the same stellar masses are systematically larger than their analogs at high-$z$ \citep{vanderWel2014}. These suggest that later-type galaxies  in low-density environments  tend to be formed with larger sizes, compared with those in dense environments. If UDGs are a tail of the high-spin  population of galaxies, one could expect that they are more numerous in  the low-density environments. On the other hand, UDGs might barely survive in the overdense regions because of strong tidal disruption, leading to a decrease of UDGs.

The detection rate of UDGs in the HCG\,95 field is higher than that in the Coma cluster, where 47 UDGs were reported over an area of 8.3 degree$^2$ by \citet{van Dokkum2015a}. We point out that the depth of survey images affects the detection rate of UDGs. Our images are as deep as those presented by \cite{FliriTrujillo2016} and \cite{Koda2015}, and deeper than the images by \cite{van Dokkum2015a}. Indeed, \cite{Koda2015} claimed that a larger number of UDGs are found in the Coma cluster with deeper imaging data \citep{Yagi2016}. In addition, the detection rate of UDGs is a strong function of the host environment \citep{vanderBurg2016,RomanTrujillo2016}. As shown in Figure~\ref{fig:fig10}, the halo mass of the Coma cluster is higher than the mass of the HCG\,95 group, and the abundance of UDGs in HCG\,95 is consistent with the halo mass-abundance correlation. Moreover, \cite{van Dokkum2017} pointed out that their estimate of the UDG abundance is incomplete because the UDGs detected in the SDSS survey were removed from their analysis (e.g., Dragonfly~X1).

We find that UDGs in the HCG\,95 field span over a wide color range, as shown in Figure~\ref{fig:fig8}. 
The majority of UDGs in our sample have a color comparable to that of red-sequence galaxies of the same stellar masses, consistent with the findings from previous studies \citep{van Dokkum2015a,vanderBurg2016}.   Still, a number of UDGs exhibit a color as blue as typical blue-cloud galaxies and H\,{\small I}-selected galaxies. There is no doubt that these blue UDGs are still forming stars. No dependence of color is found on the separation distance from the cluster center. We note that UDGs in the Pegasus~II cluster tend to be slightly bluer than those in the Pegasus~I cluster.  However, this may be due to biases in sample selection as the two subsamples of UDGs show a systematic offset in luminosity. 

The presence of blue UDGs indicates that  environmental processes quenching star formation in galaxy clusters play a less important role in the Pegasus clusters compared to Coma. \cite{Martinez-Delgado2016} reported DGSAT~I to be a UDG in PPS and argued that tidal heating effects in combination with ram-pressure stripping remove its gas content and quench star formation in DGSAT~1.   Indeed,  the lower velocity dispersion of the Pegasus~I cluster together with the lack of dense hot intracluster medium (ICM) suggest that ram pressure from hot ICM is no longer a major process for stripping gas in galaxies, and the role of interstellar medium (ISM)-ICM interactions in low-density and low-velocity dispersion environments is a possible explanation \citep{Levy2007}. The proposed mechanisms for the formation of UDGs in dense environments (e.g., rich clusters), including galaxy harassment, ram-pressure stripping and galaxy starvation, do not work for UDGs formed in low-density environments (such as poor clusters and galaxy groups).  
However, in poor clusters and galaxy groups, the formation of UDGs may be mainly affected by ISM-ICM interactions given their low-velocity dispersions. 

Interestingly, we find the UDG H95F in our sample to be a gas-rich galaxy.  This UDG has a blue $g-r$ color and strong H\,{\small I} emission detected by VLA. The blue color of H95F suggests that it is dominated by young stellar populations. H95F shows that UDGs partially overlay with H\,{\small I}-selected LSB galaxies. The H\,{\small I}-rich LSB galaxies often have a high spin and low star formation efficiency \citep{AmoriscoLoeb2016}. Our finding reveals that at least part of the UDGs originate from the same parent population as H\,{\small I}-rich LSB galaxies but appear to be extremely low in surface brightness, which might be coupled with the tail of spin at the high end. This is consistent with the theoretical interpretation by \citet{AmoriscoLoeb2016}. In such cases, star formation would not be at a sufficiently high rate to generate the strong outflows that are responsible for the termination of star formation and gas cooling at early time, which is required by the formation mechanism for UDGs seen today \citep{Janowiecki2015,Trujillo2017}. Feedback from SNe and massive stars driven gas outflows is suggested to inject energy into H\,{\small I} gas, resulting in the expansion of stellar disks and dark matter halos, and the formation of UDGs in the field that mimic LSB galaxies \citep{Di Cintio2016}. 
We argue that a tidal origin for H95F is unlikely.  Although tidal features are clearly seen in the two central galaxies of HCG\,95,  we do not find apparent tidal structures around H95F from our deep $g$-band image.  It is worth noting that PPS seems to contain about 50\% H\,{\small I}-rich galaxies \citep{Richter1982}.  The Pegasus~I cluster (and HCG\,95) appears to host more  H\,{\small I} gas-rich galaxies.  We expect that more UDGs in our sample could be detected with H\,{\small I} in future surveys of higher sensitivities. The gas-rich environment implies that the evolution of galaxies around the Pegasus clusters is still undergoing gas accretion, and the build-up of disks tends to take a longer time compared to other regions. Again, this is consistent with the idea that (at least) some of the UDGs belong to the tail of the high-spin galaxy population formed at a relatively late time.

Owing to the complexity of environmental effects, UDGs may have multiple pathways of formation and evolution. Previous studies suggested that some UDGs could be ``failed'' galaxies that are overwhelmingly dominated by dark matter \citep{van Dokkum2015a,van Dokkum2016,van Dokkum2017}. The high abundance of globular clusters found in one UDG indicates that it may be a failed galaxy with a halo mass similar to the masses of the LMC or M33 \citep{BeasleTrujillo2016,Peng2016}. Other studies revealed that UDGs are part of a dwarf galaxy population \citep{Yozin2015,AmoriscoLoeb2016,Amorisco2016,Beasley2016,BeasleTrujillo2016,Di Cintio2016}. Recently, \citet{RomanTrujillo2016} pointed  out that UDGs may be born in the field, further grow in groups, and ultimately fall into galaxy clusters by group accretion, leading to the decrease in UDG density toward dense environments.  

In addition, our results show that 26\% blue UDGs are detected in our observations. This high abundance of blue UDGs is likely associated with the environment of poor galaxy clusters residing in H\,{\small I}-rich large-scale structures. Taken together, our results indicate that the abundance of blue UDGs tends to be higher in low-density environments that are still developing, with rich gas available for slowly feeding galaxies to build up LSB disks.

\section{Summary} \label{sec:summary}

We obtained deep $g$- and $r$ images of the HCG\,95 field  with the 1\,m CNEOST.  Over an area of 4.9\,degree$^2$, we detected a sample of 89 UDG candidates that are expected to be linked with the two poor galaxy clusters Pegasus~I at $z=0.013$ and Pegasus~II at $z=0.040$. We analyzed the properties of these UDG candidates using available multiwavelength data.  We summarize our results as follows:

\begin{itemize}
\item[(1)] There are about 50 $-$ 60 true UDGs with $r_{\rm e} > 1.5$\,kpc and $\mu(g,0)>24$\,mag\,arcsec$^{-2}$ in the HCG\,95 field. These UDGs are most likely associated with the two poor galaxy clusters. This abundance of UDGs is higher than the abundance of UDGs in the field. It becomes clear that UDGs can be found ubiquitously in different environments, but a large diversity of the abundance of UDGs is seen. 

\item[(2)] 
Our UDGs are spread over a wide range in color $g-r$, covering the color regimes of both red-sequence and blue-cloud galaxies. About 23 of them are as blue as blue-cloud galaxies, suggesting that these UDGs are still forming stars. No correlation is found among UDGs between color and separation distance from the density center. Our result indicates that the environmental processes for quenching galaxy star formation appear not to be weak in the volume around HCG\,95.

\item[(3)] 
The morphologies of some UDGs appear to be irregular, the colors of these irregular UDGs are bluer than regular UDGs, like H95F. This indicates that the irregular UDGs may be connected with loose environments and have a different formation mechanism.

\item[(4)] 
Our most striking finding is the discovery that UDG H95F is a gas-rich galaxy. This is the first UDG found with solid H\,{\small I} observations by the VLA. Our finding reveals that at least some UDGs may be gas-rich galaxies and overlap the galaxies with low surface brightness galaxies detected in deep H\,{\small I} surveys. 
This supports the picture that (at least some of the) UDGs belong to the tail of the high-spin galaxy population formed at a relatively late time. 

\end{itemize}

Taken together, our results imply that the abundance of blue UDGs tends to be higher in low-density environments that are still developing, with rich gas available for slowly feeding galaxies. More efforts are needed to determine the environmental complexity and understand the formation of UDGs and their evolutionary pathways.

\acknowledgments

We are grateful to the referee for the valuable comments and suggestions that have improved our manuscript. 
This work is supported by the National Basic Research Program of China (973 Program 2013CB834900), the National Natural Science Foundation of China through grant U1331110, and the Chinese Academy of Sciences (CAS), through a grant to the CAS South America Center for Astronomy (CASSACA) in Santiago, Chile.


\clearpage
\appendix

Here we present additional technical details of our data analysis. 

Figure~\ref{fig:figA1} shows the PSF FWHM map for the $g$- and $r$-band mosaic science images. It is shown that the PSF varies over our wide-field region in the sense that it increases from the center to the edges. We use point sources to fit a third-order polynomial and build up the maps. The typical PSF FWHM is 4$\arcsec$ and 3$\arcsec$ at the center region of the $g$- and $r$-band mosaicked image, respectively.

Figure~\ref{fig:figA2} shows a comparison of our $g$- and $r$-band photometry with the SDSS for bright stars (14$< g <$18\,mag) in different annular regions of the FOV. We divide our area into four annular regions, including $R < 0\fdg37$, $0\fdg37 < R < 0\fdg74$, $0\fdg74 < R < 1\fdg11$ and $1\fdg11 < R < 1\fdg45$. Our photometry is in agreement with the photometric magnitude of the SDSS.

In Figure~\ref{fig:figA3} we compare the half-light radius of bright galaxies (15$< g <$19\,mag) in the different annular regions from the center to the edges.  We also divide our area into four annular regions, including $R < 0\fdg37$, $0\fdg37 < R < 0\fdg74$, $0\fdg74 < R < 1\fdg11$ and $1\fdg11 < R < 1\fdg45$. We use the SDSS half-light radius of bright galaxies (15$< g <$19\,mag) to calibrate our measurements. The top four panels of Figure~\ref{fig:figA3} show the comparison of the observed half-light radius without PSF correction. It is clearly seen that the PSF effect on the measurement of the half-light radius becomes larger for more compact objects with $r_{\rm SDSS} < 3\arcsec$. Some outliers are mainly affected by the contamination of scattered light and the asterism from saturation stars. 
The bottom panels show the measurements of the half-light radius corrected for PSF. The median ratio of half-light radius between the SDSS and Xuyi is 0.85, the dispersion is 0.16. This discrepancy is likely caused by the systematic effects related to pixel scale and instrumental effects. The size of the PSF ($ R_{\rm psf}$) is about $2\farcs7 - 3\farcs5$ in the $g$-band mosaicked image. We therefore use the PSF-corrected half-light radius to calculate the central surface brightness for selecting UDGs (see Section~\ref{sec:candidates}), although the measurement errors are relatively large. 

Table~\ref{tab:taba1} lists the catalog of our selected UDGs in the HCG\,95 field. We assume that 89 extremely LSB galaxies are spatially separated members of the clusters Pegasus~I and Pegasus~II cluster. Finally, we pick out 32 very LSB galaxies (with a size is smaller than 1.5\,kpc), 12 plausible UDGs, and 45 UDGs at the distance of the clusters Pegasus~I and Pegasus~II, respectively.
The properties of these candidates are shown in Table~\ref{tab:taba1}. Fortunately, 84 of the 89 very LSB galaxies are detected in deep $z\arcmin$-band images by DECaLS. The magnitudes and half-light radii of the 84 candidates are listed in Table~\ref{tab:taba1}. We use brackets to represent the lower size limits because some UDGs have very low signal-to-noise at $z\arcmin$-band image.


\begin{figure}[htb!]
 \setcounter{figure}{0} \renewcommand{\thefigure}{A.\arabic{figure}} 
 \begin{center}
  \includegraphics[trim=0mm 20mm 0mm 1mm,clip,height=0.40\textwidth]{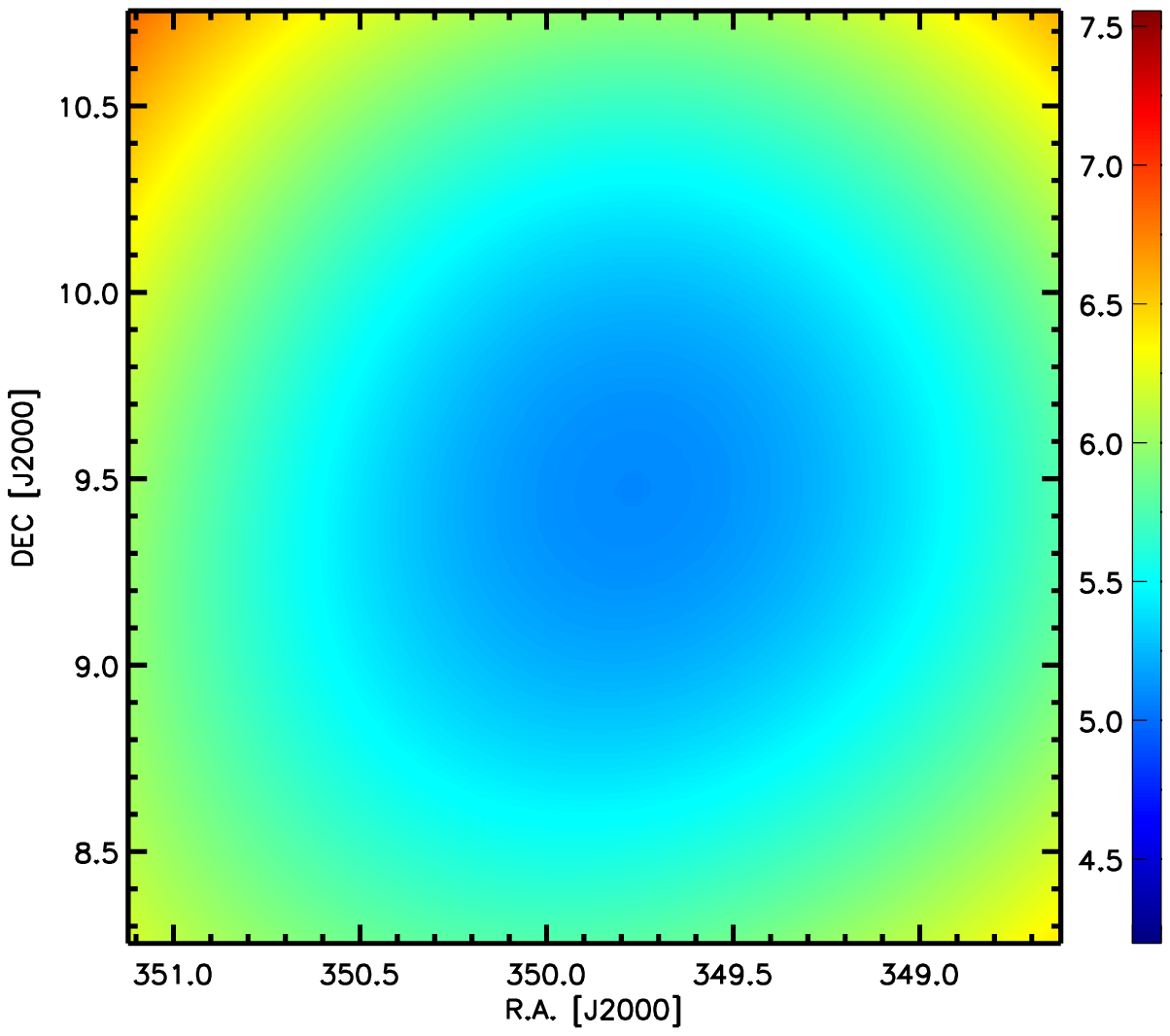}
  \includegraphics[trim=0mm 20mm 0mm 1mm,clip,height=0.40\textwidth]{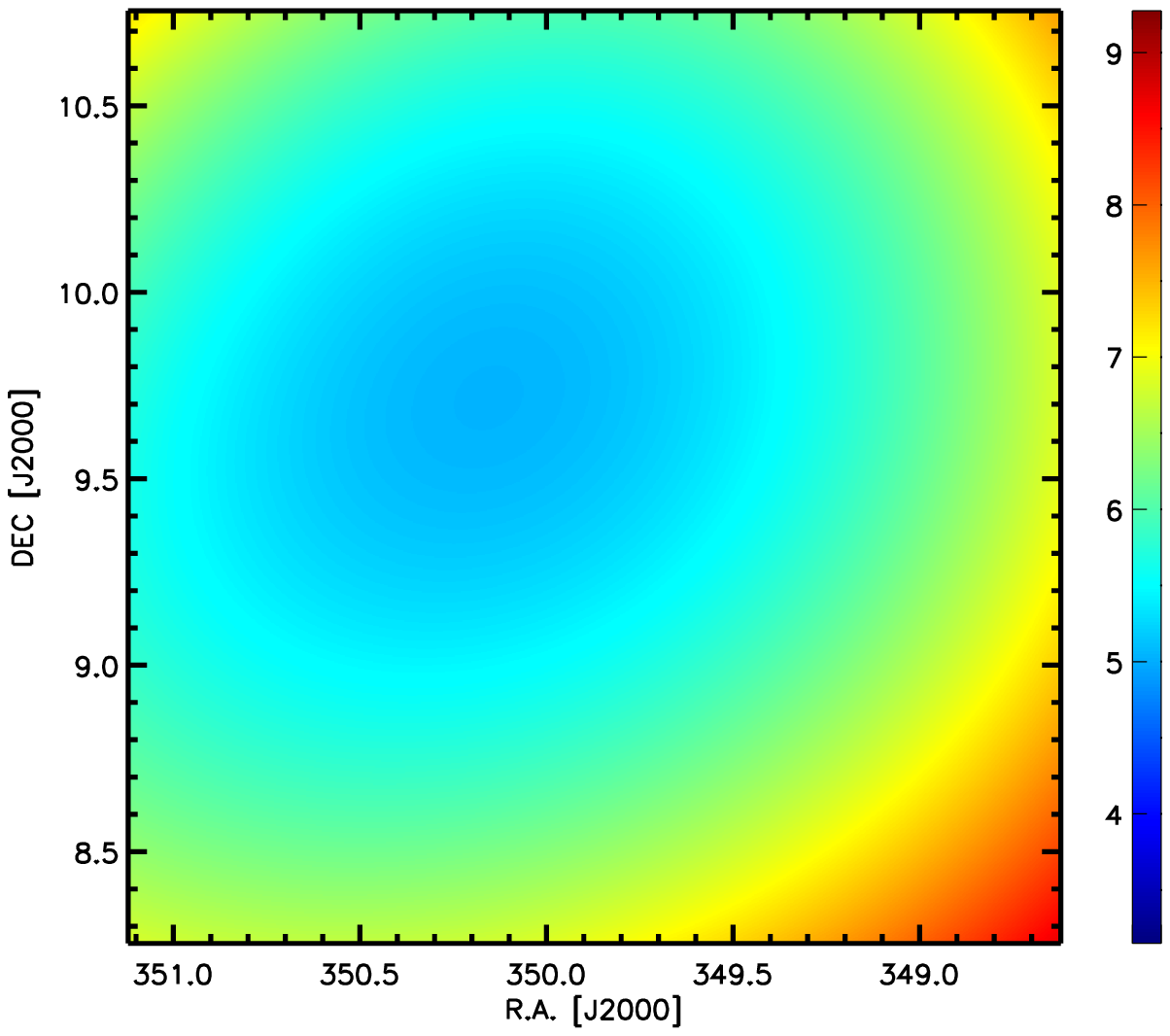}
  
  \caption{The PSF FWHM map for the $g$- and $r$-band mosaicked images. The PSF varies over our wide-field region and it increases from the center to the edges. Left: the PSF FWHM map for the $g$-band image in the $2\fdg5 \times 2\fdg5$ region. Right: the PSF FWHM map for $r$-band image in the $2\fdg5 \times 2\fdg5$ region.\label{fig:figA1}}
  \label{fig:figA1}
 \end{center}
\end{figure}

\begin{figure}[htb]
 \setcounter{figure}{1} \renewcommand{\thefigure}{A.\arabic{figure}} 

 \begin{center}
  \setlength{\abovecaptionskip}{-5pt}
  \includegraphics[trim=0mm 0mm 0mm 0mm,clip,height=0.52\textwidth]{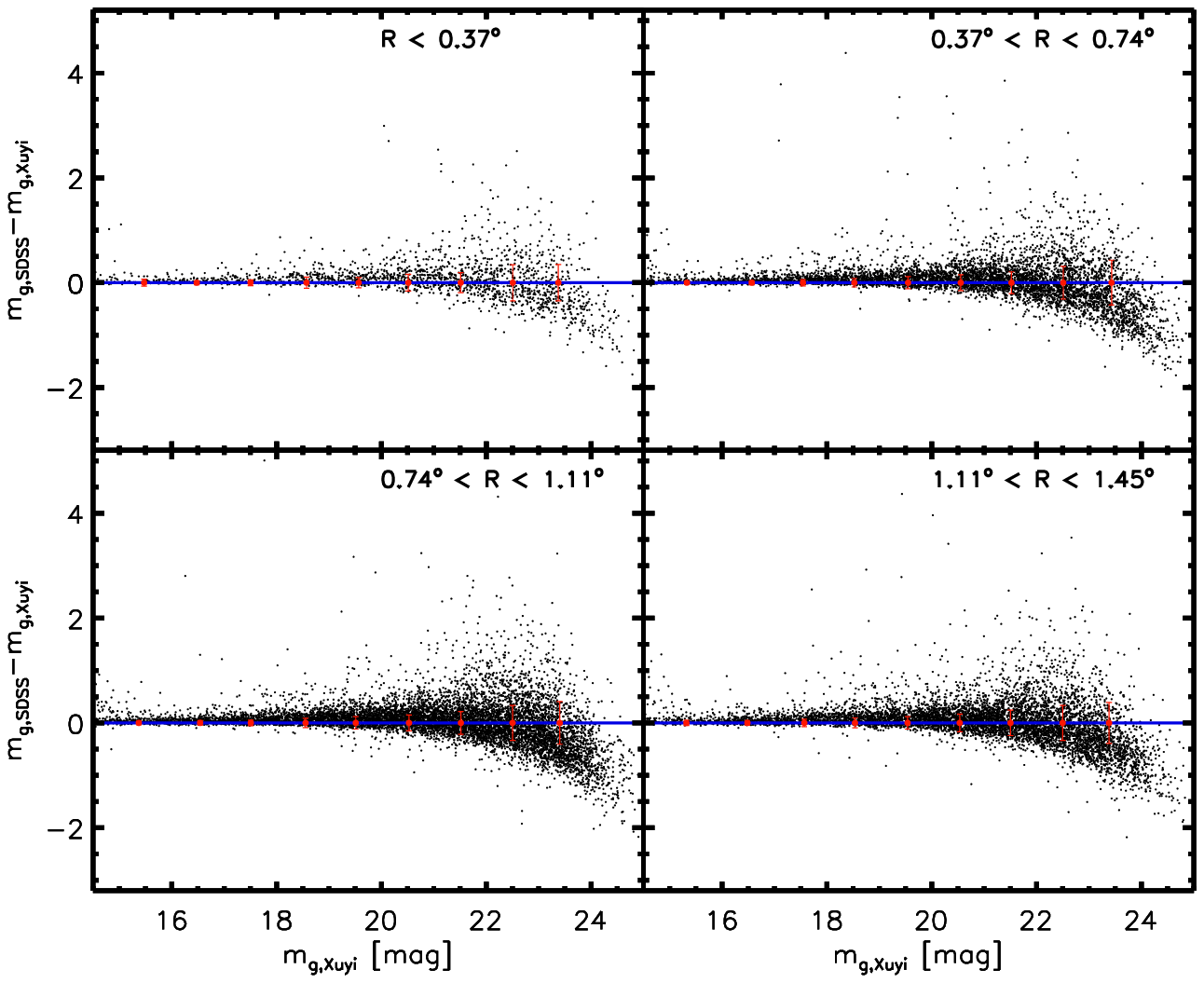}
  \includegraphics[trim=0mm 0mm 0mm 0mm,clip,height=0.52\textwidth]{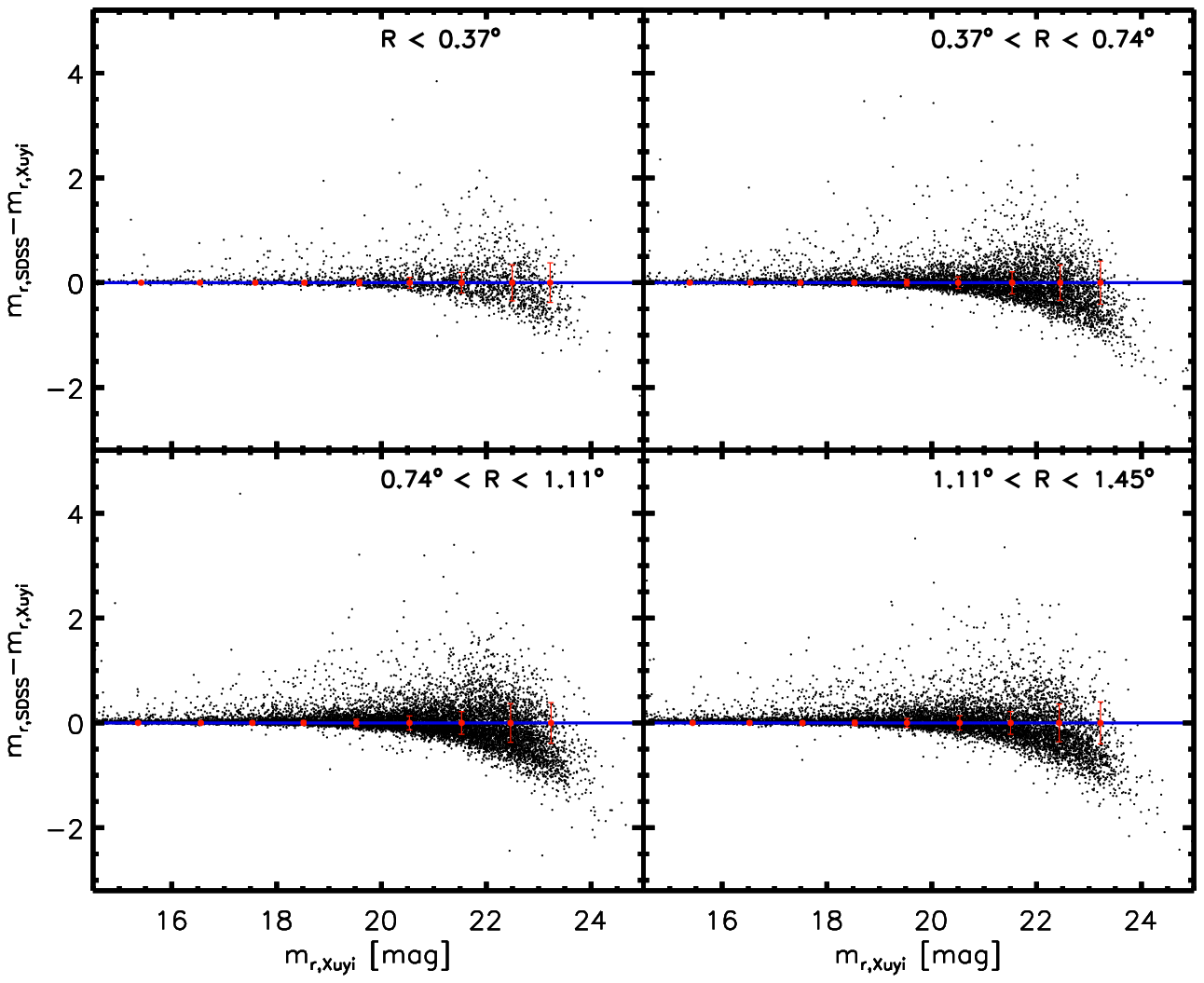}
  \caption{Comparison of our $g$- and $r$-band photometry with the SDSS for bright stars ($14< g <18$\,mag) in different annular regions. The dispersion of faint stars is larger than bright stars. Our photometry is in agreement with the photometric magnitude of the SDSS. The top panels denote the $g$-band photometry with the SDSS for bright stars, the bottom panels show the $r$-band photometry with the SDSS for bright stars.}
  \label{fig:figA2}
 \end{center}
\end{figure}

\clearpage

\begin{figure}[htb!]
 \setcounter{figure}{2} \renewcommand{\thefigure}{A.\arabic{figure}} 

 \begin{center}
   \setlength{\abovecaptionskip}{-5pt}

  \includegraphics[trim=0mm 0mm 0mm 0mm,clip,height=0.5\textwidth]{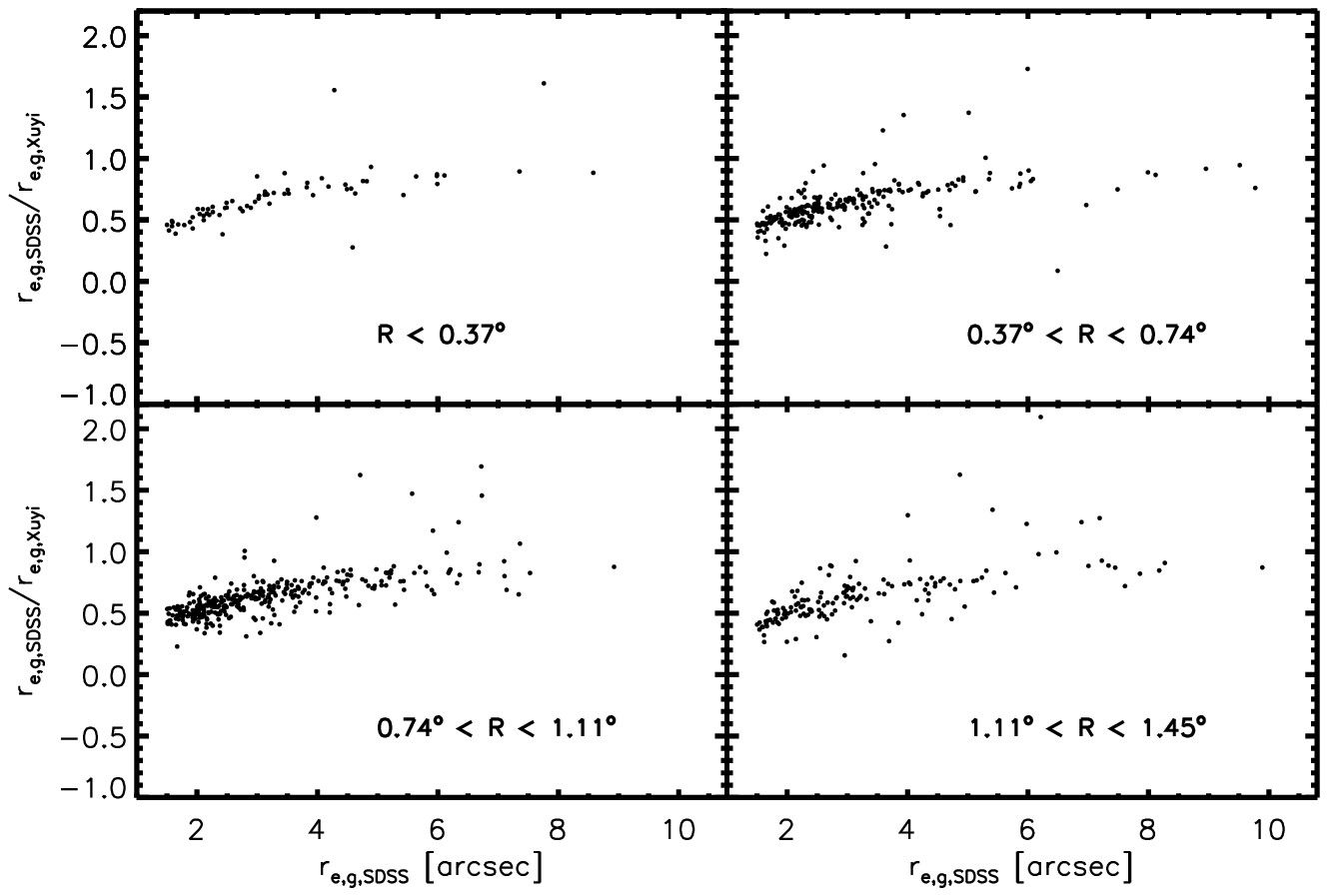}
   \includegraphics[trim=0mm 0mm 0mm 0mm,clip,height=0.5\textwidth]{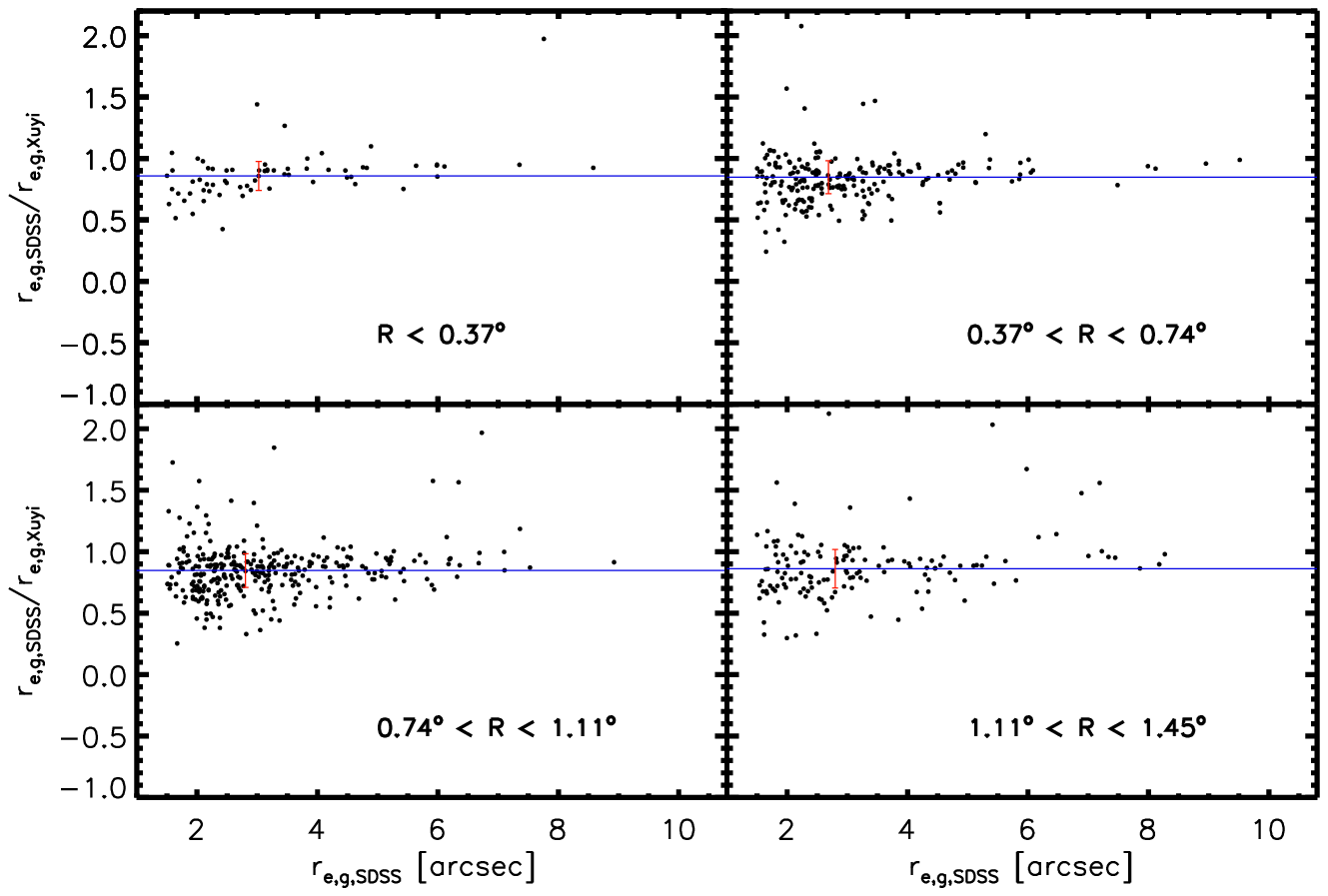}

  \caption{Comparison of the half-light radius of bright galaxies (15$< g <$19\,mag) in the different regions from the center to the edges. Upper panels: comparison of the observed half-light radius without PSF correction. Bottom panels: measurements of the half-light radius corrected for PSF.}
  \label{fig:figA3}
 \end{center}
\end{figure}

\begin{longrotatetable}
\clearpage

\begin{deluxetable*}{lllrrrrrrll}
\tabletypesize{\scriptsize}
\tablecolumns{10}
\setcounter{table}{0} \renewcommand{\thetable}{A.\arabic{table}} 
\tablecaption{Properties of UDGs in the HCG\,95 field\label{tab:taba1}}
\tablewidth{0pt}

\tablehead{
     \colhead{ID}   & \colhead{R.A.}  & \colhead{Decl.}  & \colhead{$\mu(g,0)$}    & \colhead{r$_{\rm e}$}  &\colhead{$M_{g}$} & \colhead{$b/a$} & \colhead{$g - r$} & \colhead{DECaLs($z\arcmin$)} & \colhead{DECaLs(r$_{\rm e},z\arcmin$)} \\
          &\colhead{(J2000.0)} & \colhead{(J2000.0)} &\colhead{(mag arcsec$^{-2}$)} & \colhead{(kpc)} & \colhead{(mag)} & & \colhead{(mag)}& \colhead{(mag)} & \colhead{(kpc)}  }

    \startdata
      PegI-LSBG01   & \colhead{23:19:03.6}  & \colhead{8:23:15} &\colhead{26.36}  & \colhead{1.38} & \colhead{$-$12.29} & \colhead{0.90} & \colhead{0.49} & \colhead{21.24} & \colhead{[0.49]}     \\
      PegI-LSBG02   & \colhead{23:18:35.6}  & \colhead{8:23:39} &\colhead{24.58}  & \colhead{0.97} & \colhead{$-$13.61} & \colhead{0.94} & \colhead{0.65} & \colhead{18.93} & \colhead{0.78}     \\
      PegI-LSBG03   & \colhead{23:19:43.8}  & \colhead{8:26:23} &\colhead{24.00}  & \colhead{0.93} & \colhead{$-$14.03} & \colhead{0.80} & \colhead{0.48} & \colhead{18.57} & \colhead{0.78}      \\
      PegI-LSBG04   & \colhead{23:17:49.9}  & \colhead{8:28:05} &\colhead{25.13}  & \colhead{1.32} & \colhead{$-$13.60} & \colhead{0.82} & \colhead{0.67} & \colhead{19.18} & \colhead{0.80}      \\
      PegI-LSBG05   & \colhead{23:20:46.9}  & \colhead{8:28:29} &\colhead{24.88}  & \colhead{1.32} & \colhead{$-$13.69} & \colhead{0.72} & \colhead{0.75} & \colhead{19.46} & \colhead{0.72}      \\
      PegI-LSBG06   & \colhead{23:18:28.1}  & \colhead{8:28:58} &\colhead{26.45}  & \colhead{1.33} & \colhead{$-$12.46} & \colhead{0.97} & \colhead{0.53} & \colhead{21.77} & \colhead{[0.36]}      \\
      PegI-LSBG07   & \colhead{23:19:11.9}  & \colhead{8:33:32} &\colhead{25.09}  & \colhead{0.91} & \colhead{$-$12.78} & \colhead{0.79} & \colhead{0.64} & \colhead{20.39} & \colhead{[0.50]}     \\
      PegI-LSBG08   & \colhead{23:20:37.4}  & \colhead{8:35:26} &\colhead{24.29}  & \colhead{1.21} & \colhead{$-$13.80} & \colhead{0.55} & \colhead{0.53} & \colhead{19.04} & \colhead{0.64}      \\
      PegI-LSBG09   & \colhead{23:21:45.5}  & \colhead{8:37:41} &\colhead{25.02}  & \colhead{0.86} & \colhead{$-$12.25} & \colhead{0.51} & \colhead{0.74} & \colhead{20.49} & \colhead{[0.32]}      \\
      PegI-LSBG10   & \colhead{23:20:13.1}  & \colhead{8:39:31} &\colhead{24.58}  & \colhead{1.24} & \colhead{$-$13.59} & \colhead{0.56} & \colhead{0.72} & \colhead{18.97} & \colhead{0.71}      \\
      PegI-LSBG11   & \colhead{23:21:10.6}  & \colhead{8:39:45} &\colhead{24.10}  & \colhead{1.06} & \colhead{$-$13.52} & \colhead{0.46} & \colhead{0.60} & \colhead{19.11} & \colhead{0.63}      \\
      PegI-LSBG12   & \colhead{23:22:08.4}  & \colhead{8:41:60} &\colhead{24.27}  & \colhead{0.56} & \colhead{$-$12.26} & \colhead{0.61} & \colhead{0.39} & \colhead{21.22} & \colhead{[0.27]}   \\
      PegI-LSBG13   & \colhead{23:18:22.2}  & \colhead{8:42:53} &\colhead{25.41}  & \colhead{0.65} & \colhead{$-$11.73} & \colhead{0.81} & \colhead{0.67} & \colhead{21.89} & \colhead{[0.19]}     \\
      PegI-LSBG14   & \colhead{23:19:47.0}  & \colhead{8:42:33} &\colhead{24.62}  & \colhead{1.01} & \colhead{$-$13.55} & \colhead{0.84} & \colhead{0.67} & \colhead{19.00} & \colhead{0.74}      \\
      PegI-LSBG15   & \colhead{23:21:22.9}  & \colhead{8:43:11} &\colhead{26.32}  & \colhead{1.05} & \colhead{$-$11.99} & \colhead{0.89} & \colhead{0.96} & \colhead{22.05} & \colhead{[0.19]}      \\
      PegI-LSBG16   & \colhead{23:22:55.7}  & \colhead{8:44:46} &\colhead{25.74}  & \colhead{1.07} & \colhead{$-$11.99} & \colhead{0.51} & \colhead{0.41} & \colhead{21.52}   & \colhead{[0.25]}    \\
      PegI-LSBG17   & \colhead{23:22:09.1}  & \colhead{8:48:49} &\colhead{25.56}  & \colhead{1.01} & \colhead{$-$12.01} & \colhead{0.49} & \colhead{0.93} & \colhead{21.15} & \colhead{[0.36]}      \\ 
      PegI-LSBG18   & \colhead{23:23:24.5}  & \colhead{8:50:49} &\colhead{25.31}  & \colhead{0.86} & \colhead{$-$12.21} & \colhead{0.60} & \colhead{0.57} & \colhead{21.34} & \colhead{[0.28]}     \\
      PegI-LSBG19   & \colhead{23:20:13.5}  & \colhead{8:54:31} &\colhead{24.25}  & \colhead{0.95} & \colhead{$-$13.66} & \colhead{0.53} & \colhead{0.75} & \colhead{18.94}   & \colhead{0.71}   \\
      PegI-LSBG20   & \colhead{23:22:58.0}  & \colhead{8:54:43} &\colhead{24.00}  & \colhead{0.89} & \colhead{$-$13.88} & \colhead{0.78} & \colhead{0.64} & \colhead{18.79}   & \colhead{0.69}    \\
      PegI-LSBG21   & \colhead{23:22:10.7}  & \colhead{8:56:52} &\colhead{25.83}  & \colhead{1.00} & \colhead{$-$12.04} & \colhead{0.65} & \colhead{0.82} & \colhead{21.23} & \colhead{[0.27]}      \\
      PegI-LSBG22   & \colhead{23:23:59.3}  & \colhead{8:57:03} &\colhead{25.76}  & \colhead{0.95} & \colhead{$-$12.04} & \colhead{0.68} & \colhead{0.89} & \colhead{20.91} & \colhead{[0.23]}      \\
      PegI-LSBG23   & \colhead{23:23:00.8}  & \colhead{8:59:56} &\colhead{24.92}  & \colhead{1.01} & \colhead{$-$13.11} & \colhead{0.75} & \colhead{0.67} & \colhead{19.87} & \colhead{0.74}      \\
      PegI-LSBG24   & \colhead{23:21:36.2}  & \colhead{9:03:41} &\colhead{24.14}  & \colhead{0.88} & \colhead{$-$13.47} & \colhead{0.66} & \colhead{0.76} & \colhead{19.60} & \colhead{0.61}      \\
      PegI-LSBG25   & \colhead{23:22:28.3}  & \colhead{9:04:19} &\colhead{24.05}  & \colhead{0.94} & \colhead{$-$13.74} & \colhead{0.54} & \colhead{0.65} & \colhead{19.35}   & \colhead{0.55}    \\
      PegI-LSBG26   & \colhead{23:20:41.8}  & \colhead{9:09:02} &\colhead{24.55}  & \colhead{0.97} & \colhead{$-$12.92} & \colhead{0.50} & \colhead{0.75} & \colhead{20.37} & \colhead{[0.45]}      \\
      PegI-LSBG27   & \colhead{23:24:25.3}  & \colhead{9:12:22} &\colhead{24.02}  & \colhead{0.98} & \colhead{$-$14.14} & \colhead{0.88} & \colhead{0.34} & \colhead{19.00} & \colhead{0.74}     \\
      PegI-LSBG28   & \colhead{23:22:44.7}  & \colhead{9:19:52} &\colhead{24.17}  & \colhead{0.82} & \colhead{$-$13.64} & \colhead{0.92} & \colhead{0.27} & \colhead{19.71} & \colhead{0.51}      \\
      PegI-LSBG29   & \colhead{23:21:39.5}  & \colhead{9:21:36} &\colhead{24.33}  & \colhead{1.04} & \colhead{$-$13.92} & \colhead{0.86} & \colhead{0.42} & \colhead{18.98} & \colhead{0.75}   \\
      PegI-LSBG30   & \colhead{23:23:36.9}  & \colhead{9:37:48} &\colhead{25.84}  & \colhead{0.76} & \colhead{$-$11.62} & \colhead{0.80} & \colhead{0.38} & \colhead{21.61} & \colhead{[0.28]}      \\
      PegI-LSBG31   & \colhead{23:23:46.8}  & \colhead{9:43:59} &\colhead{24.11}  & \colhead{0.94} & \colhead{$-$13.24} & \colhead{0.46} & \colhead{0.72} & \colhead{19.55} & \colhead{0.46}      \\
      PegI-LSBG32   & \colhead{23:24:04.5}  & \colhead{9:44:43} &\colhead{25.20}  & \colhead{1.15} & \colhead{$-$13.32} & \colhead{0.90} & \colhead{0.41} & \colhead{20.31} & \colhead{[0.43]}      \\
     \hline
      PegI-UDG01   & \colhead{23:20:04.9}  & \colhead{8:21:29} &\colhead{24.43}  & \colhead{1.65} & \colhead{$-$14.31} & \colhead{0.53} & \colhead{0.70} & \colhead{18.21} & \colhead{0.88}   \\
      PegI-UDG02   & \colhead{23:21:07.3}  & \colhead{8:23:44} &\colhead{25.02}  & \colhead{1.93} & \colhead{$-$14.11} & \colhead{0.56} & \colhead{0.69} & \colhead{18.93} & \colhead{0.89}      \\
      PegI-UDG03   & \colhead{23:21:57.3}  & \colhead{8:26:27} &\colhead{25.51}  & \colhead{2.75} & \colhead{$-$14.85} & \colhead{0.85} & \colhead{0.69} & \colhead{20.15} & \colhead{[0.49]}      \\
      PegI-UDG04   & \colhead{23:21:34.4}  & \colhead{8:28:39} &\colhead{25.28}  & \colhead{2.49} & \colhead{$-$14.66} & \colhead{0.71} & \colhead{0.83} & \colhead{18.72} & \colhead{1.16}      \\  
      PegI-UDG05   & \colhead{23:21:34.6}  & \colhead{8:29:26} &\colhead{24.67}  & \colhead{2.44} & \colhead{$-$15.29} & \colhead{0.77} & \colhead{0.76} & \colhead{17.40} & \colhead{1.68}   \\
      PegI-UDG06   & \colhead{23:20:20.8}  & \colhead{8:33:05} &\colhead{24.38}  & \colhead{1.61} & \colhead{$-$14.80} & \colhead{0.85} & \colhead{0.71} & \colhead{17.99} & \colhead{0.97}     \\
      PegI-UDG07   & \colhead{23:19:57.8}  & \colhead{8:33:19} &\colhead{24.47}  & \colhead{1.55} & \colhead{$-$14.39} & \colhead{0.68} & \colhead{0.57} & \colhead{18.64}   & \colhead{0.80}    \\
      PegI-UDG08   & \colhead{23:18:52.5}  & \colhead{8:40:47} &\colhead{24.08}  & \colhead{1.88} & \colhead{$-$14.96} & \colhead{0.55} & \colhead{0.58} & \colhead{17.93} & \colhead{1.17}      \\
      PegI-UDG09   & \colhead{23:19:39.5}  & \colhead{8:42:32} &\colhead{25.45}  & \colhead{1.51} & \colhead{$-$13.13} & \colhead{0.69} & \colhead{0.80} & \colhead{20.40} & \colhead{[0.61]}     \\
      PegI-UDG10   & \colhead{23:21:01.4}  & \colhead{8:50:02} &\colhead{26.19}  & \colhead{1.86} & \colhead{$-$13.20} & \colhead{0.75} & \colhead{0.63} & \colhead{21.09} & \colhead{[0.53]}     \\
      PegI-UDG11   & \colhead{23:21:41.1}  & \colhead{9:07:12} &\colhead{24.33}  & \colhead{1.99} & \colhead{$-$15.21} & \colhead{0.77} & \colhead{0.52} & \colhead{17.51} & \colhead{1.51}      \\
      PegI-UDG12   & \colhead{23:23:59.5}  & \colhead{9:51:18} &\colhead{24.55}  & \colhead{2.33} & \colhead{$-$15.19} & \colhead{0.67} & \colhead{0.58} & \colhead{18.36} & \colhead{1.02}      \\
     \hline
      PegII-UDG01   & \colhead{23:17:07.0}  & \colhead{8:39:06} &\colhead{25.88}  & \colhead{2.92} & \colhead{$-$14.27} & \colhead{0.59} & \colhead{0.23} & \colhead{20.25} & \colhead{[1.20]}      \\
      PegII-UDG02   & \colhead{23:16:13.6}  & \colhead{8:44:13} &\colhead{24.31}  & \colhead{1.70} & \colhead{$-$15.19} & \colhead{0.94} & \colhead{0.73} & \colhead{20.12} & \colhead{[1.44]}      \\
      PegII-UDG03   & \colhead{23:16:04.3}  & \colhead{8:48:15} &\colhead{24.57}  & \colhead{1.93} & \colhead{$-$15.04} & \colhead{0.79} & \colhead{0.15} & \colhead{20.93} & \colhead{[0.91]}      \\
      PegII-UDG04   & \colhead{23:16:08.6}  & \colhead{8:53:56} &\colhead{25.29}  & \colhead{3.44} & \colhead{$-$15.28} & \colhead{0.66} & \colhead{0.63} & \colhead{19.88} & \colhead{1.92}      \\
      PegII-UDG05   & \colhead{23:18:09.2}  & \colhead{8:53:30} &\colhead{24.02}  & \colhead{2.06} & \colhead{$-$15.34} & \colhead{0.43} & \colhead{0.58} & \colhead{20.42} & \colhead{[1.11]}      \\
      PegII-UDG06   & \colhead{23:17:20.7}  & \colhead{8:59:08} &\colhead{25.18}  & \colhead{2.26} & \colhead{$-$14.67} & \colhead{0.78} & \colhead{0.40} & \colhead{21.24}   & \colhead{[0.66]}   \\
      PegII-UDG07   & \colhead{23:15:17.5}  & \colhead{8:59:35} &\colhead{24.00}  & \colhead{3.14} & \colhead{$-$16.29} & \colhead{0.56} & \colhead{0.40} & \colhead{19.14}   & \colhead{1.90}   \\
      PegII-UDG08   & \colhead{23:16:18.1}  & \colhead{9:01:14} &\colhead{24.07}  & \colhead{4.20} & \colhead{$-$16.94} & \colhead{0.60} & \colhead{0.47} & \colhead{18.85} & \colhead{2.47}      \\
      PegII-UDG09   & \colhead{23:14:57.6}  & \colhead{9:04:59} &\colhead{25.57}  & \colhead{4.23} & \colhead{$-$15.47} & \colhead{0.54} & \colhead{0.74} & \colhead{20.40} & \colhead{[1.11]}      \\
      PegII-UDG10   & \colhead{23:20:11.7}  & \colhead{9:18:27} &\colhead{24.0}   & \colhead{3.64} & \colhead{$-$16.95} & \colhead{0.76} & \colhead{0.40} & \colhead{18.50} & \colhead{2.59}      \\
      PegII-UDG11   & \colhead{23:15:28.6}  & \colhead{9:18:45} &\colhead{24.83}  & \colhead{4.40} & \colhead{$-$16.38} & \colhead{0.71} & \colhead{0.57} & \colhead{\nodata}   & \colhead{\nodata}    \\
      PegII-UDG12   & \colhead{23:18:02.2}  & \colhead{9:20:18} &\colhead{24.01}  & \colhead{1.61} & \colhead{$-$15.06} & \colhead{0.71} & \colhead{0.57} & \colhead{20.63}   & \colhead{[1.09]}    \\
      PegII-UDG13   & \colhead{23:18:56.0}  & \colhead{9:20:18} &\colhead{25.08}  & \colhead{2.97} & \colhead{$-$15.22} & \colhead{0.65} & \colhead{0.37} & \colhead{21.09} & \colhead{[1.49]}      \\
      PegII-UDG14   & \colhead{23:18:58.2}  & \colhead{9:21:41} &\colhead{24.0}  & \colhead{2.49} & \colhead{$-$16.20} & \colhead{0.78} & \colhead{0.70} & \colhead{19.01} & \colhead{1.99}      \\
      PegII-UDG15   & \colhead{23:18:10.2}  & \colhead{9:27:13} &\colhead{24.09}  & \colhead{3.33} & \colhead{$-$16.23} & \colhead{0.54} & \colhead{0.72} & \colhead{19.07}   & \colhead{1.40}   \\
      PegII-UDG16   & \colhead{23:20:36.8}  & \colhead{9:30:02} &\colhead{25.04}  & \colhead{2.36} & \colhead{$-$14.92} & \colhead{0.75} & \colhead{0.69} & \colhead{20.15} & \colhead{[1.56]}      \\
      PegII-UDG17   & \colhead{23:17:44.9}  & \colhead{9:29:38} &\colhead{24.00}  & \colhead{3.79} & \colhead{$-$17.07} & \colhead{0.72} & \colhead{0.48} & \colhead{18.44}   & \colhead{3.05}    \\
      PegII-UDG18   & \colhead{23:17:34.4}  & \colhead{9:30:16} &\colhead{25.36}  & \colhead{2.74} & \colhead{$-$15.02} & \colhead{0.82} & \colhead{0.40} & \colhead{20.92}   & \colhead{[1.42]}    \\
      PegII-UDG19   & \colhead{23:19:21.5}  & \colhead{9:30:14} &\colhead{24.12}  & \colhead{2.94} & \colhead{$-$16.27} & \colhead{0.88} & \colhead{0.40} & \colhead{19.02} & \colhead{2.26}      \\
      PegII-UDG20   & \colhead{23:18:32.5}  & \colhead{9:31:54} &\colhead{24.04}  & \colhead{3.34} & \colhead{$-$16.57} & \colhead{0.58} & \colhead{0.56} & \colhead{19.63} & \colhead{1.58}      \\
      PegII-UDG21   & \colhead{23:20:22.6}  & \colhead{9:32:58} &\colhead{24.48}  & \colhead{2.28} & \colhead{$-$15.14} & \colhead{0.58} & \colhead{0.44} & \colhead{20.40} & \colhead{[1.60]}      \\
      PegII-UDG22   & \colhead{23:17:38.8}  & \colhead{9:33:26} &\colhead{24.00}  & \colhead{2.31} & \colhead{$-$15.90} & \colhead{0.74} & \colhead{0.55} & \colhead{19.27}   & \colhead{2.21} \\
 PegII-UDG23\tablenotemark{a}   & \colhead{23:19:29.0}  & \colhead{9:33:31} &\colhead{24.05}  & \colhead{3.41} & \colhead{$-$16.69} & \colhead{0.74} & \colhead{0.28} & \colhead{20.10} & \colhead{[1.25]}     \\
      PegII-UDG24   & \colhead{23:18:10.1}  & \colhead{9:35:14} &\colhead{24.53}  & \colhead{2.85} & \colhead{$-$15.84} & \colhead{0.76} & \colhead{0.66} & \colhead{19.41}   & \colhead{1.80}   \\
      PegII-UDG25   & \colhead{23:22:19.5}  & \colhead{9:37:07} &\colhead{24.43}  & \colhead{5.66} & \colhead{$-$17.15} & \colhead{0.52} & \colhead{0.36} & \colhead{18.65} & \colhead{3.23}      \\
      PegII-UDG26   & \colhead{23:14:41.4}  & \colhead{9:39:07} &\colhead{24.15}  & \colhead{1.74} & \colhead{$-$14.82} & \colhead{0.55} & \colhead{0.74} & \colhead{21.57} & \colhead{[0.85]}      \\
      PegII-UDG27   & \colhead{23:14:50.4}  & \colhead{9:39:02} &\colhead{25.32}  & \colhead{3.36} & \colhead{$-$14.82} & \colhead{0.60} & \colhead{0.68} & \colhead{22.22} & \colhead{[0.73]}      \\
      PegII-UDG28   & \colhead{23:20:36.5}  & \colhead{9:40:16} &\colhead{25.56}  & \colhead{2.98} & \colhead{$-$14.53} & \colhead{0.53} & \colhead{0.83} & \colhead{\nodata}   & \colhead{\nodata} \\
      PegII-UDG29   & \colhead{23:21:31.6}  & \colhead{9:42:09} &\colhead{24.32}  & \colhead{4.11} & \colhead{$-$16.56} & \colhead{0.58} & \colhead{0.70} & \colhead{19.06} & \colhead{1.84}     \\
      PegII-UDG30   & \colhead{23:21:02.0}  & \colhead{9:45:17} &\colhead{24.34}  & \colhead{2.97} & \colhead{$-$15.86} & \colhead{0.58} & \colhead{0.76} & \colhead{20.35} & \colhead{[1.58]}      \\
      PegII-UDG31   & \colhead{23:18:04.1}  & \colhead{9:50:29} &\colhead{24.38}  & \colhead{3.15} & \colhead{$-$16.39} & \colhead{0.89} & \colhead{0.56} & \colhead{19.28}   & \colhead{1.87} \\
      PegII-UDG32   & \colhead{23:17:45.4}  & \colhead{9:50:43} &\colhead{24.22}  & \colhead{1.73} & \colhead{$-$14.86} & \colhead{0.65} & \colhead{0.99} & \colhead{21.18}   & \colhead{[1.10]}   \\
      PegII-UDG33   & \colhead{23:17:40.1}  & \colhead{9:50:51} &\colhead{24.16}  & \colhead{2.87} & \colhead{$-$16.12} & \colhead{0.68} & \colhead{0.80} & \colhead{19.22}   & \colhead{1.79}   \\
      PegII-UDG34   & \colhead{23:17:27.7}  & \colhead{9:52:12} &\colhead{24.13}  & \colhead{2.67} & \colhead{$-$15.92} & \colhead{0.64} & \colhead{0.52} & \colhead{19.69} & \colhead{1.77}      \\
      PegII-UDG35   & \colhead{23:17:08.8}  & \colhead{9:53:55} &\colhead{24.00}  & \colhead{3.50} & \colhead{$-$17.11} & \colhead{0.98} & \colhead{0.58} & \colhead{\nodata}   & \colhead{\nodata}    \\
      PegII-UDG36   & \colhead{23:22:24.7}  & \colhead{9:54:15} &\colhead{24.80}  & \colhead{4.04} & \colhead{$-$16.28} & \colhead{0.84} & \colhead{0.59} & \colhead{19.24} & \colhead{1.66}      \\
      PegII-UDG37   & \colhead{23:22:08.2}  & \colhead{9:54:47} &\colhead{24.20}  & \colhead{4.11} & \colhead{$-$17.14} & \colhead{0.89} & \colhead{0.39} & \colhead{18.97} & \colhead{2.29}      \\
      PegII-UDG38   & \colhead{23:18:56.3}  & \colhead{9:55:18}  &\colhead{24.04}  & \colhead{3.37} & \colhead{$-$16.37} & \colhead{0.55} & \colhead{0.46} & \colhead{19.20} & \colhead{1.92}   \\
      PegII-UDG39   & \colhead{23:23:59.5}  & \colhead{10:11:01} &\colhead{24.07}  & \colhead{3.30} & \colhead{$-$16.49} & \colhead{0.67} & \colhead{0.47} & \colhead{18.79} & \colhead{1.95}      \\
      PegII-UDG40   & \colhead{23:20:10.9}  & \colhead{10:16:12} &\colhead{25.22}  & \colhead{3.85} & \colhead{$-$15.68} & \colhead{0.67} & \colhead{0.53} & \colhead{20.07} & \colhead{[1.42]}      \\
      PegII-UDG41   & \colhead{23:21:35.6}  & \colhead{10:17:56} &\colhead{24.50}  & \colhead{3.86} & \colhead{$-$16.22} & \colhead{0.56} & \colhead{0.62} & \colhead{\nodata}   & \colhead{\nodata}    \\
      PegII-UDG42   & \colhead{23:22:00.3}  & \colhead{10:21:12} &\colhead{24.38}  & \colhead{2.73} & \colhead{$-$16.16} & \colhead{0.96} & \colhead{0.46} & \colhead{19.00} & \colhead{2.02}      \\
      PegII-UDG43   & \colhead{23:20:28.5}  & \colhead{10:28:12} &\colhead{24.44}  & \colhead{2.94} & \colhead{$-$15.90} & \colhead{0.69} & \colhead{0.64} & \colhead{19.81} & \colhead{1.65}      \\
      PegII-UDG44   & \colhead{23:18:10.6}  & \colhead{10:33:50} &\colhead{25.01}  & \colhead{3.04} & \colhead{$-$15.37} & \colhead{0.66} & \colhead{0.60} & \colhead{\nodata}   & \colhead{\nodata}   \\
      PegII-UDG45   & \colhead{23:18:43.5}  & \colhead{10:35:40} &\colhead{24.14}  & \colhead{2.06} & \colhead{$-$15.38} & \colhead{0.65} & \colhead{0.39} & \colhead{20.32} & \colhead{[1.56]}      \\
     \enddata
     \label{tab:tableA1}
\tablenotetext{a}{H95F = PegII-UDG25}
\end{deluxetable*}
\end{longrotatetable}


\begin{thebibliography}{a}
\bibitem[Abell et al.(1989)]{Abell1989} Abell, G.~O., Corwin, H.~G., Jr., \& Olowin, R.~P.\ 1989, \apjs, 70, 1 
\bibitem[Abraham \& van Dokkum(2014)]{Abraham2014} Abraham, R.~G., \& van Dokkum, P.~G.\ 2014, \pasp, 126, 55 
\bibitem[Amorisco \& Loeb et al.(2016)]{AmoriscoLoeb2016} Amorisco, N.~C. \& Loeb, A., \ 2016, \mnras, 459, L51
\bibitem[Amorisco et al.(2016)]{Amorisco2016} Amorisco, N.~C., Monachesi, A., \& White, S.~D.~M.\ 2016, arXiv:1610.01595 
\bibitem[Barnes et al.(1999)]{Barnes1999} Barnes, D.~G., Webster, R.~L., Schmidt, R.~W., \& Hughes, A.\ 1999, \mnras, 309, 641 
\bibitem[Beasley et al.(2016)]{Beasley2016} Beasley, M.~A., Romanowsky, A.~J., Pota, V., et al.\ 2016, \apjl, 819, L20 
\bibitem[Beasley \& Trujillo(2016)]{BeasleTrujillo2016} Beasley, M.~A., \& Trujillo, I.\ 2016, \apj, 830, 23 
\bibitem[Bell et al.(2003)]{Bell2003} Bell, E.~F., McIntosh, D.~H., Katz, N., \& Weinberg, M.~D.\ 2003, \apjs, 149, 289 
\bibitem[Bellazzini et al.(2017)]{Bellazzini2017} Bellazzini, M., Belokurov, V., Magrini, L., et al.\ 2017, \mnras, 467, 3751 
\bibitem[Bertin \& Arnouts(1996)]{BertinArnouts1996} Bertin, E., \& Arnouts, S.\ 1996, \aaps, 117, 393 
\bibitem[Bertin et al.(2002)]{Bertin2002} Bertin, E., Mellier, Y., Radovich, M., et al. 2002, in ASP Conf. Ser. 281, Astronomical Data Analysis Software and Systems XI, ed. D. A. Bohlender, D. Durand, \& T. H. Handley (San Francisco, CA: ASP), 228
\bibitem[Bertin et al.(2006)]{Bertin2006} Bertin, E. 2006, in ASP Conf. Ser. 351, Astronomical Data Analysis Software and Systems XV, ed. C. Gabriel, C. Arviset, D. Ponz, \& E. Solano (San Francisco, CA: ASP), 112 
\bibitem[Blanton et al.(2006)]{Blanton2006} Blanton, M.~R., Eisenstein, D., Hogg, D.~W., \& Zehavi, I.\ 2006, \apj, 645, 977 
\bibitem[Blum et al.(2016)]{Blum2016} Blum, R.~D., Burleigh, K., Dey, A., et al.\ 2016, AAS Meeting, 228, 317.01 
\bibitem[Bothun et al.(1991)]{Bothun1991} Bothun, G.~D., Impey, C.~D., \& Malin, D.~F.\ 1991, \apj, 376, 404 
\bibitem[Bouch{\'e} et al.(2015)]{Bouche2015} Bouch{\'e}, N., Carfantan, H., Schroetter, I., Michel-Dansac, L., \& Contini, T.\ 2015, \aj, 150, 92 
\bibitem[Brodie et al.(2011)]{Brodie2011} Brodie, J.~P., Romanowsky, A.~J., Strader, J., \& Forbes, D.~A.\ 2011, \aj, 142, 199 
\bibitem[Bruzual \& Charlot(2003)]{Bruzual2003} Bruzual, G., \& Charlot, S.\ 2003, \mnras, 344, 1000 
\bibitem[Canizares et al.(1986)]{Canizares1986} Canizares, C.~R., Donahue, M.~E., Trinchieri, G., Stewart, G.~C., \& McGlynn, T.~A.\ 1986, \apj, 304, 312 
\bibitem[Ceccarelli et al.(2012)]{Ceccarelli2012} Ceccarelli, L., Herrera-Camus, R., Lambas, D.~G., Galaz, G., \& Padilla, N.~D.\ 2012, \mnras, 426, L6
\bibitem[Chincarini et al.(1976)]{Chincarini1976} Chincarini, G., \& Rood, H.~J.\ 1976, \pasp, 88, 388 
\bibitem[Di Cintio et al.(2017)]{Di Cintio2016} Di Cintio, A., Brook, C.~B., Dutton, A.~A., et al.\ 2017, \mnras, 466, L1 
\bibitem[Da Rocha et al.(2005)]{DaRocha2005} Da Rocha, C. \& Mendes de Oliveira, C. \ 2005, \mnras, 364, 1069
\bibitem[Falco et al.(1999)]{Falco1999} Falco, E.~E., Kurtz, M.~J., Geller, M.~J., et al.\ 1999, \pasp, 111, 438 
\bibitem[Fliri \& Trujillo(2016)]{FliriTrujillo2016} Fliri, J., \& Trujillo, I.\ 2016, \mnras, 456, 1359 
\bibitem[Freyhammer et al.(2001)]{Freyhammer2001} Freyhammer, L.~M., Andersen, M.~I., Arentoft, T., Sterken, C., \& N{\o}rregaard, P.\ 2001, Experimental Astronomy, 12, 147 
\bibitem[Fukugita et al.(1996)]{Fukugita1996} Fukugita, M., Ichikawa, T., Gunn, J.~E., et al.\ 1996, \aj, 111, 1748 
\bibitem[Gavazzi et al.(2005)]{Gavazzi2005} Gavazzi, G., Donati, A., Cucciati, O., et al.\ 2005, \aap, 430, 411 
\bibitem[Geller et al.(2012)]{Geller2012} Geller, M.~J., Diaferio, A., Kurtz, M.~J., Dell'Antonio, I.~P., \& Fabricant, D.~G.\ 2012, \aj, 143, 102
\bibitem[Giovanelli et al.(2005)]{Giovanelli2005} Giovanelli, R., Haynes, M.~P., Kent, B.~R., et al.\ 2005, \aj, 130, 2598 
\bibitem[Girardi et al.(1998)]{Girardi1998} Girardi, M., Giuricin, G., Mardirossian, F., Mezzetti, M., \& Boschin, W.\ 1998, \apj, 505, 74 
\bibitem[Graham \& Driver(2005)]{Graham2005} Graham, A.~W., \& Driver, S.~P.\ 2005, \pasa, 22, 118 
\bibitem[Haynes et al.(2011)]{Haynes2011} Haynes, M.~P., Giovanelli, R., Martin, A.~M., et al.\ 2011, \aj, 142, 170 
\bibitem[Hickson et al.(1992)]{Hickson1992} Hickson, P., Mendes de Oliveira, C., Huchra, J.~P., \& Palumbo, G.~G.\ 1992, \apj, 399, 353
\bibitem[Huchtmeier et al.(2000)]{Huchtmeier2000} Huchtmeier, W.~K., Verdes-Montenegro, L., Yun, M., del Olmo, A., \& Perea, J.\ 2000, in IAU Colloq.~174: Small Galaxy Groups, ed.M.J.Valtonen \& C.Flynn (San Francisco,CA: ASP),154
\bibitem[Iglesias-P{\'a}ramo \& V{\'{\i}}lchez(1997)]{Iglesias-Paramo1997} Iglesias-P{\'a}ramo, J., \& V{\'{\i}}lchez, J.~M.\ 1997, \apjl, 489, L13 
\bibitem[Iglesias-P{\'a}ramo \& V{\'{\i}}lchez(1998)]{Iglesias-Paramo1998} Iglesias-P{\'a}ramo, J., \& V{\'{\i}}lchez, J.~M.\ 1998, \aj, 115, 1791 
\bibitem[Iglesias-P{\'a}ramo \& V{\'{\i}}lchez(2001)]{Iglesias-Paramo2001} Iglesias-P{\'a}ramo, J., \& V{\'{\i}}lchez, J.~M.\ 2001, \apj, 550, 204 
\bibitem[Impey et al.(1988)]{Impey1988} Impey, C., Bothun, G., \& Malin, D.\ 1988, \apj, 330, 634 
\bibitem[Impey et al.(1996)]{Impey1996} Impey, C.~D., Sprayberry, D., Irwin, M.~J., \& Bothun, G.~D.\ 1996, \apjs, 105, 209 
\bibitem[Impey \& Bothun(1997)]{Impey1997} Impey, C., \& Bothun, G.\ 1997, \araa, 35, 267
\bibitem[Impey et al.(2001)]{Impey2001} Impey, C., Burkholder, V., \& Sprayberry, D.\ 2001, \aj, 122, 2341
\bibitem[Janowiecki et al.(2015)]{Janowiecki2015} Janowiecki, S., Leisman, L., J{\'o}zsa, G., et al.\ 2015, \apj, 801, 96 
\bibitem[Janssens et al.(2017)]{Janssens2017} Janssens, S., Abraham, R., Brodie, J., et al.\ 2017, \apjl, 839, L17 
\bibitem[Koch et al.(2017)]{Koch2016} Koch, A., Black, C.~S., Rich, R.~M., et al.\ 2017, Astronomische Nachrichten, 338, 503 
\bibitem[Koda et al.(2015)]{Koda2015} Koda, J., Yagi, M., Yamanoi, H., \& Komiyama, Y.\ 2015, \apjl, 807, L2
\bibitem[Leisman et al.(2017)]{Leisman2017} Leisman, L., Haynes, M.~P., Janowiecki, S., et al.\ 2017, \apj, 842, 133 
\bibitem[Levy et al.(2007)]{Levy2007} Levy, L., Rose, J.~A., van Gorkom, J.~H., \& Chaboyer, B.\ 2007, \aj, 133, 1104 
\bibitem[Mart{\'i}nez-Delgado et al.(2016)]{Martinez-Delgado2016} Mart{\'{\i}}nez-Delgado, D., L{\"a}sker, R., Sharina, M., et al.\ 2016, \aj, 151, 96
\bibitem[Merritt et al.(2016)]{Merritt2016} Merritt, A., van Dokkum, P., Danieli, S., et al.\ 2016, \apj, 833, 168 
\bibitem[Mihos et al.(2015)]{Mihos2015} Mihos, J.~C., Durrell, P.~R., Ferrarese, L., et al.\ 2015, \apjl, 809, L21 
\bibitem[Mu{\~n}oz et al.(2015)]{Munoz2015} Mu{\~n}oz, R.~P., Eigenthaler, P., Puzia, T.~H., et al.\ 2015, \apjl, 813, L15 
\bibitem[Noonan(1981)]{Noonan1981} Noonan, T.~W.\ 1981, \apjs, 45, 613 
\bibitem[O'Neil et al.(1997)]{O'Neil1997} O'Neil, K., Bothun, G.~D., \& Cornell, M.~E.\ 1997, \aj, 113, 1212 
\bibitem[Ordenes-Brice{\~n}o et al.(2016)]{Ordenes-Briceno2016} Ordenes-Brice{\~n}o, Y., Taylor, M.~A., Puzia, T.~H., et al.\ 2016, \mnras, 463, 1284 
\bibitem[Peng et al.(2002)]{peng2002} Peng, C.~Y., Ho, L.~C., Impey, C.~D., \& Rix, H.-W.\ 2002, \aj, 124, 266 
\bibitem[Peng \& Lim(2016)]{Peng2016} Peng, E.~W., \& Lim, S.\ 2016, \apjl, 822, L31 
\bibitem[Ponman et al.(1996)]{Ponman1996} Ponman, T.~J., Bourner, P.~D.~J., Ebeling, H., et al.\ 1996, \mnras, 283, 690 
\bibitem[Randall et al.(2009)]{Randall2009} Randall, S.~W., Jones, C., Kraft, R., Forman, W.~R., \& O'Sullivan, E.\ 2009, \apj, 696, 1431 
\bibitem[Richter \& Huchtmeier(1982)]{Richter1982} Richter, O.-G., \& Huchtmeier, W.~K.\ 1982, \aap, 109, 155 
\bibitem[Rodrigue et al.(1995)]{Rodrigue1995} Rodrigue, M., Schultz, A., Thompson, J., et al.\ 1995, \aj, 109, 2362 
\bibitem[Rom{\'a}n \& Trujillo(2017a)]{Roman2017} Rom{\'a}n, J., \& Trujillo, I.\ 2017, \mnras, 468, 703 
\bibitem[Rom{\'a}n \& Trujillo(2017b)]{RomanTrujillo2016} Rom{\'a}n, J., \& Trujillo, I.\ 2017, \mnras, 468, 4039 
\bibitem[Rood \& Struble(1994)]{Rood1994} Rood, H.~J., \& Struble, M.~F.\ 1994, \pasp, 106, 413 
\bibitem[Smith Castelli et al.(2016)]{Smith2016} Smith Castelli, A.~V., Faifer, F.~R., \& Escudero, C.~G.\ 2016, \aap, 596, A23 
\bibitem[Taylor et al.(2011)]{Taylor2011} Taylor, E.~N., Hopkins, A.~M., Baldry, I.~K., et al.\ 2011, \mnras, 418, 1587 
\bibitem[Teimoorinia et al.(2017)]{Teimoorinia2017} Teimoorinia, H., Ellison, S.~L., \& Patton, D.~R.\ 2017, \mnras, 464, 3796 
\bibitem[Thomas et al.(2005)]{Thomas05} Thomas, D., Maraston, C., Bender, R., \& Mendes de Oliveira, C.\ 2005, \apj, 621, 673 
\bibitem[Trujillo et al.(2017)]{Trujillo2017} Trujillo, I., Roman, J., Filho, M., \& S{\'a}nchez Almeida, J.\ 2017, \apj, 836, 191 
\bibitem[Valtchanov et al.(1999)]{Valtchanov1999} Valtchanov, I., Kalinkov, M., \& Kuneva, I.\ 1999, in Proc. Int. Conf. on Numerical Astrophysics, ed. S.M. Miyama, K.Tomisaka, \& T. Hanawa (Boston, MA: Kluwer Academic),59 
\bibitem[van der Burg et al.(2016)]{vanderBurg2016} van der Burg, R.~F.~J., Muzzin, A., \& Hoekstra, H.\ 2016, \aap, 590, A20
\bibitem[van der Wel et al.(2014)]{vanderWel2014} van der Wel, A., Chang, Y.-Y., Bell, E.~F., et al.\ 2014, \apjl, 792, L6 
\bibitem[van Dokkum et al.(2016)]{van Dokkum2016} van Dokkum, P., Abraham, R., Brodie, J., et al.\ 2016, \apjl, 828, L6 
\bibitem[van Dokkum et al.(2017)]{van Dokkum2017} van Dokkum, P., Abraham, R., Romanowsky, A.~J., et al.\ 2017, \apjl, 844, L11 
\bibitem[van Dokkum et al.(2015a)]{van Dokkum2015a} van Dokkum, P.~G., Abraham, R., Merritt, A., et al.\ 2015, \apjl, 798, L45
\bibitem[van Dokkum et al.(2015b)]{van Dokkum2015b} van Dokkum, P.~G., Romanowsky, A.~J., Abraham, R., et al.\ 2015, \apjl, 804, L26 
\bibitem[Verdes-Montenegro et al.(1997)]{Verdes-Montenegro1997} Verdes-Montenegro, L., Yun, M., Perea, J., \& del Olmo, A.\ 1997, in ASP Conf. 117, Dark and Visible Matter in Galaxies and Cosmological Implications, ed. M. Persic \& P. Salucci (San Francisco, CA: ASP),530 
\bibitem[Yagi et al.(2016)]{Yagi2016} Yagi, M., Koda, J., Komiyama, Y., \& Yamanoi, H.\ 2016, \apjs, 225, 11 
\bibitem[Yozin \& Bekki(2015)]{Yozin2015} Yozin, C., \& Bekki, K.\ 2015, \mnras, 452, 937 
\bibitem[Zhang et al.(2013)]{Zhang2013} Zhang, H.-H., Liu, X.-W., Yuan, H.-B., et al.\ 2013, Research in Astronomy and Astrophysics, 13, 490 
\bibitem[Zhang et al.(2014)]{Zhang2014} Zhang, H.-H., Liu, X.-W., Yuan, H.-B., et al.\ 2014, Research in Astronomy and Astrophysics, 14, 456
\bibitem[Zhong et al.(2008)]{Zhong2008} Zhong, G.~H., Liang, Y.~C., Liu, F.~S., et al.\ 2008, \mnras, 391, 986 
\end{thebibliography}
\end{document}